\documentclass[twocolumn,times,twocolappendix]{aastex631}
\pdfoutput=1

\usepackage{xspace}
\usepackage{booktabs}
\usepackage{longtable}
\usepackage{threeparttablex}

\newcommand{\msun}{{\rm M}_{\odot}}

\newcommand{\zsun}{{\rm Z}_{\odot}}

\newcommand{\kms}{{\rm km\,s^{-1}}}

\newcommand{\mesa}{\mbox{\uppercase{Mesa}}\xspace}
\newcommand{\eg}{e.g.\@\xspace}
\newcommand{\cf}{c.f.\@\xspace}
\newcommand{\ie}{i.e.\@\xspace}
\newcommand{\casea}{Case~A\@\xspace}
\newcommand{\caseb}{Case~B\@\xspace}
\newcommand{\casebe}{Case~Be\@\xspace}
\newcommand{\casebl}{Case~Bl\@\xspace}
\newcommand{\casec}{Case~C\@\xspace}

\newcommand{\casebh}{Case-B\@\xspace}

\newcommand{\casech}{Case-C\@\xspace}
\newcommand{\caseabh}{Case-AB\@\xspace}

\newcommand{\BSS}{BSS\@\xspace}
\newcommand{\BSSs}{BSSs\@\xspace}
\newcommand{\bhl}{BH\textsubscript{L}\@\xspace}
\newcommand{\bhh}{BH\textsubscript{H}\@\xspace}
\shorttitle{Bimodal black-hole mass distribution}
\shortauthors{Schneider et al.}
\graphicspath{{./}{figures/}}

\begin{document}

\title{Bimodal black-hole mass distribution and chirp masses of binary black-hole mergers}

\correspondingauthor{Fabian R.~N.~Schneider}
\email{fabian.schneider@h-its.org}

\author[0000-0002-5965-1022]{Fabian R.~N.~Schneider}
\affiliation{Heidelberger Institut f{\"u}r Theoretische Studien, Schloss-Wolfsbrunnenweg 35, 69118 Heidelberg, Germany}
\affiliation{Astronomisches Rechen-Institut, Zentrum f{\"u}r Astronomie der Universit{\"a}t Heidelberg, M{\"o}nchhofstr.\ 12-14, 69120 Heidelberg, Germany}

\author[0000-0002-8338-9677]{Philipp~Podsiadlowski}
\affiliation{University of Oxford, St Edmund Hall, Oxford, OX1 4AR, United Kingdom}
\affiliation{Heidelberger Institut f{\"u}r Theoretische Studien, Schloss-Wolfsbrunnenweg 35, 69118 Heidelberg, Germany}

\author[0000-0003-1009-5691]{Eva~Laplace}
\affiliation{Heidelberger Institut f{\"u}r Theoretische Studien, Schloss-Wolfsbrunnenweg 35, 69118 Heidelberg, Germany}



\received{2 May 2023}
\revised{16 May 2023}
\accepted{17 May 2023}
\submitjournal{ApJ Letters}

\begin{abstract} 
In binary black-hole mergers from isolated binary-star evolution, both black holes are from progenitor stars that have lost their hydrogen-rich envelopes by binary mass transfer. Envelope stripping is known to affect the pre-supernova core structures of such binary-stripped stars and thereby their final fates and compact remnant masses. In this paper, we show that binary-stripped stars give rise to a bimodal black-hole mass spectrum with characteristic black-hole masses of about $9\,\msun$ and $16\,\msun$ across a large range of metallicities. The bimodality is linked to carbon and neon burning becoming neutrino-dominated, which results in interior structures that are difficult to explode and likely lead to black hole formation. The characteristic black-hole masses from binary-stripped stars have corresponding features in the chirp-mass distribution of binary black-hole mergers: peaks at about $8$ and $14\,\msun$, and a dearth in between these masses. Current gravitational-wave observations of binary black-hole mergers show evidence for a gap at $10\text{--}12\,\msun$ and peaks at $8$ and $14\,\msun$ in the chirp-mass distribution. These features are in agreement with our models of binary-stripped stars. In the future, they may be used to constrain the physics of late stellar evolution and supernova explosions, and may even help measure the cosmological expansion of the Universe.
\end{abstract}

\keywords{Stellar evolution (1599) --- Multiple star evolution  (2153) --- Stellar remnants (1627) --- Black holes (162) --- Neutron stars (1108) --- Gravitational wave sources (677)}


\section{\label{sec:intro}Introduction}

Gravitational-wave (GW) astronomy has opened up a new window to the Universe and led to unprecedented insights \citep{2016PhRvL.116f1102A, 2016PhRvL.116x1103A, 2017PhRvL.119p1101A, 2017ApJ...848L..12A}. With almost 100 GW detections of binary neutron-star (BNS), binary black-hole (BBH) and neutron star--black hole (NS-BH) mergers \citep{2019PhRvX...9c1040A, 2021PhRvX..11b1053A, 2021arXiv211103606T}, the mass distribution of stellar-mass BHs is being revealed across cosmic time. This will help to better understand many aspects relevant to the formation of BHs such as supernova (SN) explosion physics, and pre-SN evolution of massive single and binary stars. Previously, only $\mathcal{O}(10)$ stellar-mass BHs were known, mostly residing in Galactic X-ray binaries \citep{2017hsn..book.1499C} with Cygnus X-1 hosting the most massive BH of ${\approx}\,21\,\msun$ \citep{2021Sci...371.1046M}.

The chirp mass of GW merger events,
\begin{equation}
    \mathcal{M} = \frac{(m_1 m_2)^{3/5}}{(m_1+m_2)^{1/5}},
    \label{eq:chirp-mass}
\end{equation}
with component masses $m_1$ and $m_2$, is one of the most accurately known GW observables \citep{2021ApJ...913L..19T}. Since the observing run O3a of the Advanced LIGO and Advanced Virgo instruments, hints of a gap in the chirp-mass distribution at $10\text{--}12\,\msun$ and peaks at $8$, $14$, $27$ and $45\,\msun$ have been reported \citep{2021ApJ...913L..19T}. They are corroborated by measurements in O3b \citep{2021arXiv211103634T, 2022ApJ...928..155T} and have complementary features in the inferred BH mass distribution at about $9$, $16$, $30$ and $57\,\msun$ \citep{2021ApJ...913L..19T, 2021arXiv211103634T, 2022ApJ...928..155T, 2023ApJ...946...16E}. Given the current number of BBH merger detections, the peaks at $9$ and $30\,\msun$ seem robust whereas the significance of features around $\approx16\,\msun$ is less clear \citep{2018ApJ...856..173T, 2022PhRvD.105l3014S, 2022arXiv220712409W, 2023ApJ...946...16E, 2023arXiv230100834F}.

\begin{figure}
  \centering
  \includegraphics[width=0.90\linewidth]{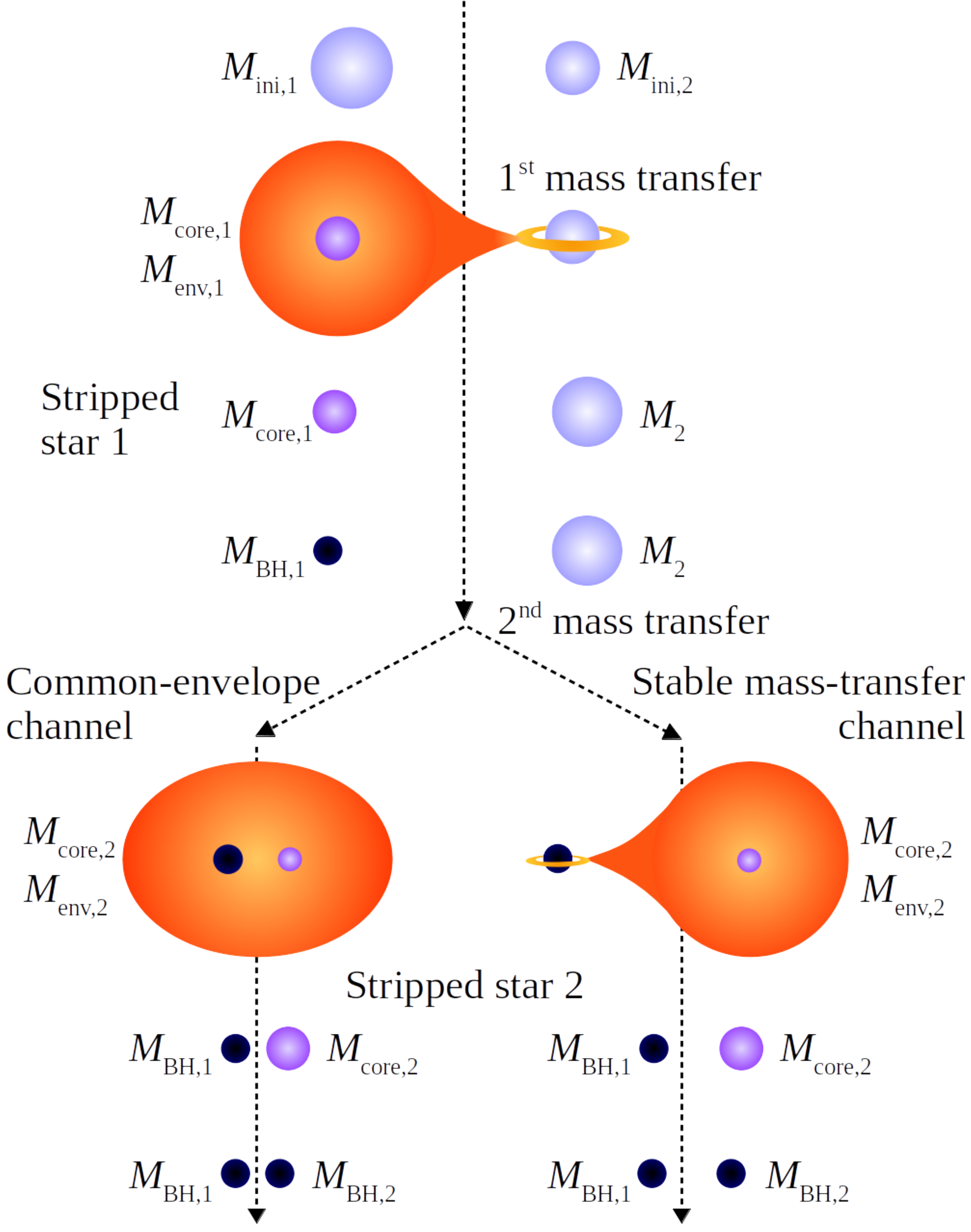}
  \caption{\label{fig:schematic-binary-evolution}Schematic evolution of isolated binary stars of initial masses $M_\mathrm{ini,1}$ and $M_\mathrm{ini,2}$ forming BBH mergers via the common-envelope and stable mass-transfer channels (not to scale). Both the first ($M_\mathrm{BH,1}$) and second ($M_\mathrm{BH,2}$) formed BHs are from \BSSs. $M_\mathrm{core}$ and $M_\mathrm{env}$ denote the core and envelope masses of the two stars, respectively. Figure courtesy of Friedrich R{\"o}pke after original cartoons by Thomas Tauris; inspired by figure~1 in \citet{2022ApJ...931...17V}.}
\end{figure}

BBH mergers are theoretically expected to form through isolated binary evolution and dynamically in dense stellar environments \citep{2021hgwa.bookE..16M, 2022hgwa.bookE..15K, 2022LRR....25....1M}. In the former channel\footnote{We do not consider chemically-homogenous evolution because it leads to BBH mergers with chirp masses ${\gtrsim}\,20\,\msun$ that are beyond the masses of interest in this work \citep{2016A&A...588A..50M, 2016MNRAS.458.2634M, 2020MNRAS.499.5941D}.} (Fig.~\ref{fig:schematic-binary-evolution}), the first- and second-born BH both originate from stars that have been stripped of their envelopes by stable binary mass transfer or a common-envelope phase (called binary-stripped stars, \BSSs, from hereon). Dynamically-formed BBH mergers additionally invoke BHs from single stars and repeated BH mergers.

Envelope removal in \BSSs is known to affect the core structures of stars such that they form NSs and BHs of different masses compared to single stars \citep{1996ApJ...457..834T, 1999A&A...350..148W, 2001NewA....6..457B, 2004ApJ...612.1044P, 2019ApJ...878...49W, 2020ApJ...890...51E, 2021A&A...645A...5S, 2021A&A...656A..58L}. In this paper, we compute the birth distributions of BH masses from \BSSs and single stars at different metallicities. We show that the peaks around the chirp-mass gap and the corresponding peaks in the BH-mass distribution are a natural outcome of envelope stripping in binary stars.

\section{\label{sec:methods}Methods}

We use published models of single and \BSSs at solar metallicity \citep{2021A&A...645A...5S}, $Z=\zsun=0.0142$, and newly compute models at $\zsun/10$ following the methodology of \citet{2021A&A...645A...5S}. We employ the stellar evolution code \mesa \citep[revision 10398,][]{2011ApJS..192....3P, 2013ApJS..208....4P, 2015ApJS..220...15P, 2018ApJS..234...34P, 2019ApJS..243...10P} and evolve models from the beginning of core-hydrogen burning until an iron core forms that collapses at velocities of ${>}\,950\,\kms$. The pre-SN stellar structures are then used as input to the neutrino-driven, semi-analytic SN code of \citet{2016MNRAS.460..742M} to determine whether stars explode in SNe or collapse to BHs. We apply calibrations as in \citet{2021A&A...645A...5S}, assume that the entire stellar mass falls into the BH, and use a maximum NS mass of $2\,\msun$. The initial masses of stars are chosen to avoid (pulsational) pair-instability SNe (for further details and all stellar model data see Appendix~\ref{app:stellar-models}). 

The pre-SN core structures of \BSSs depend on the timing of mass transfer \citep{2021A&A...645A...5S}. We consider the removal of hydrogen-rich stellar envelopes during core hydrogen burning (\casea), between the end of core hydrogen and core helium burning (\caseb), and after core helium exhaustion (\casec). \caseb is further divided into early and late for donors with radiative and convective envelopes, respectively. Because \casea/B BSSs, and single stars/\casech BSSs result in similar pre-SN core structures \citep{2021A&A...645A...5S}, we simplify our figures and only show results for single stars and \casebh BSSs.

\section{\label{sec:pre-sn-structure-and-remnant-masses}Pre-supernova stellar structures}

\begin{figure*}
    \centering
    \includegraphics{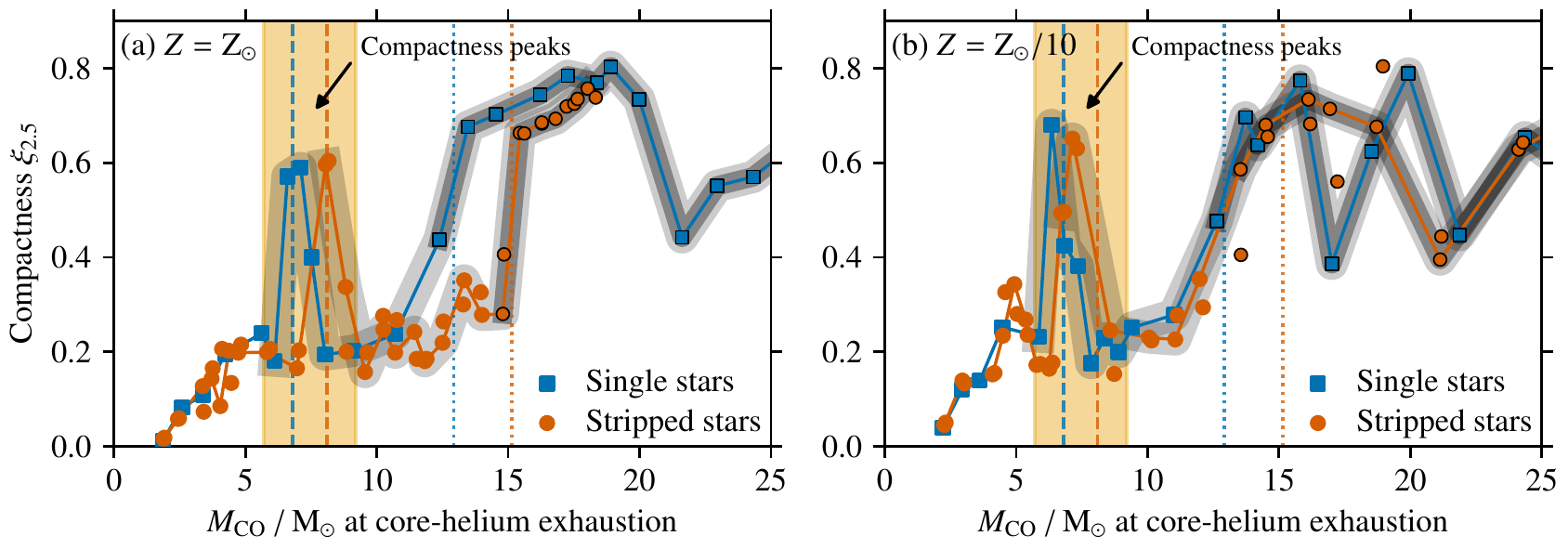}
    \caption{\label{fig:compactness}Compactness $\xi_\mathrm{2.5}$ as a function of CO-core mass $M_\mathrm{CO}$ for (a) the $Z=\zsun$ and (b) the $\zsun/10$ models. The orange-shaded region indicates the compactness peak and vertical dashed lines show its position for the $\zsun$ single (blue) and BSS (red) models. The vertical dotted lines show the corresponding second major increase in compactness of the $\zsun$ models. In \BSSs, early \casebh systems are connected by solid lines. Light-grey and dark-grey shading indicate radiative, \ie neutrino-dominated, core carbon and core neon burning, respectively, where the latter is additionally marked by black edges around the data points.} 
\end{figure*}

The delayed, neutrino-driven SN mechanism is the leading idea on how the collapse of the core of a massive star is reversed and leads to a supernova explosion \citep[\eg][]{1966PhRv..141.1232M, 1966ApJ...143..626C, 1967CaJPh..45.1621A, 2002ApJ...574L..65F, 2005ARNPS..55..467M, 2007PhR...442...38J, 2008ApJ...678.1207I, 2014ApJ...786...83T, 2016PASA...33...48M, 2016ARNPS..66..341J, 2018ApJ...865...81O, 2018JPhG...45j4001O, 2020MNRAS.491.2715B, 2021Natur.589...29B}. For the explodability of stars within this framework, an often-used proxy is the compactness of their cores at core collapse\footnote{For example, in magneto-rotational and other SN explosion mechanisms, the compactness parameter may not be a useful proxy for explodability.},
\begin{equation}
    \xi_M = \frac{M/\msun}{R(M)/1000\,\mathrm{km}},
    \label{eq:xi}
\end{equation}
where $R(M)$ is the radius at mass coordinate $M$, usually taken as $M=2.5\,\msun$ \citep{2011ApJ...730...70O}. Low compactness $\xi_\mathrm{2.5}$ suggests that stars explode while high values indicate unsuccessful explosions and collapse to BHs \citep{2011ApJ...730...70O, 2012ApJ...757...69U, 2014ApJ...783...10S, 2016ApJ...818..124E, 2016ApJ...821...38S, 2016MNRAS.460..742M, 2020ApJ...890...43C, 2020MNRAS.499.2803P, 2021A&A...645A...5S}.

The compactness of massive stars and hence their final fates are mainly set by the carbon-oxygen (CO) core mass $M_\mathrm{CO}$ and the central carbon mass fraction $X_\mathrm{C}$ \emph{at the end of core helium burning} \citep{2020ApJ...890...43C, 2020MNRAS.499.2803P, 2021A&A...645A...5S}. Stellar evolution beyond helium burning is driven by thermal neutrino losses that depend mostly on the density and temperature in the stellar cores, which are set by $M_\mathrm{CO}$. Nuclear energy generation of the subsequent carbon burning additionally depends on the amount of available fuel, \ie $X_\mathrm{C}$.

In helium burning, three $\alpha$-particles fuse into carbon and, once carbon is available, $\alpha$-capture via the $^{12}\mathrm{C}(\alpha,\gamma)^{16}\mathrm{O}$ nuclear reaction produces oxygen and consumes carbon. The number of available $\alpha$-particles further depends on the growth of the convective core during helium burning which is closely linked to shell hydrogen burning. In stars with more massive cores, $X_\mathrm{C}$ is smaller such that carbon burning eventually becomes neutrino-dominated and turns radiative at a certain $M_\mathrm{CO}$. Enhanced core contraction then leads to high compactness \citep{2014ApJ...783...10S, 2020ApJ...890...43C, 2020MNRAS.499.2803P, 2021A&A...645A...5S}. At somewhat higher $M_\mathrm{CO}$, neon and oxygen burning occur earlier, decreasing the compactness such that a compactness peak forms \citep[see below;][]{Laplace+2023}. For even higher $M_\mathrm{CO}$, also neon burning becomes neutrino-dominated, again sharply increasing the compactness \citep{2014ApJ...783...10S, 2020ApJ...890...43C, 2020MNRAS.499.2803P, 2021A&A...645A...5S}.

At $\zsun$, this leads to compactness peaks in single stars and \BSSs at $M_\mathrm{CO}\approx 7\,\msun$ and $8\,\msun$, respectively, where carbon burning becomes neutrino-dominated, and an increase of $\xi_\mathrm{2.5}$ at $M_\mathrm{CO}\approx 13\,\msun$ and $15\,\msun$, respectively, where also neon burning becomes neutrino-dominated (Fig.~\ref{fig:compactness}a; see also \citealt{2021A&A...645A...5S}). The $M_\mathrm{CO}$-shifts are caused by a larger $X_\mathrm{C}$ in \BSSs compared to single stars, because the convective helium-burning cores of \BSSs grow less in mass as a consequence of having lost the hydrogen-rich envelopes by binary mass transfer \citep{1996ApJ...463..297B, 2001NewA....6..457B, 2004ApJ...612.1044P, 2019ApJ...878...49W, 2020ApJ...890...43C, 2020MNRAS.492.2578S, 2020MNRAS.499.2803P, 2021A&A...645A...5S, 2021A&A...656A..58L, Laplace+2023}. At $\zsun/10$, we find the same qualitative $\xi_\mathrm{2.5}\text{--}M_\mathrm{CO}$ relations (Fig.~\ref{fig:compactness}b).

\begin{figure*}
    \centering
    \includegraphics{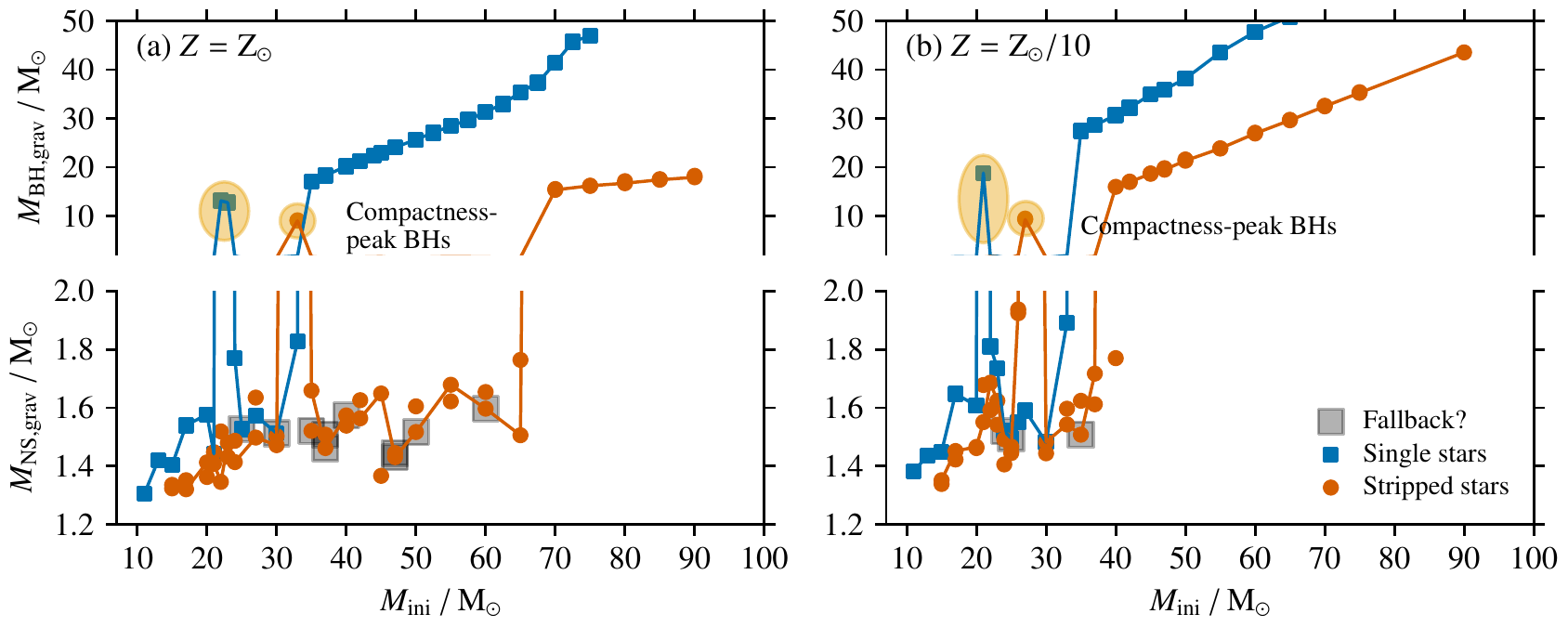}
    \caption{\label{fig:compact-remnant-masses}Gravitational NS and BH masses ($M_\mathrm{NS,grav}$ and $M_\mathrm{BH,grav}$, respectively) as a function of initial stellar mass for the single stars and \BSSs at (a) $Z=\zsun$ and (b) $\zsun/10$. As in Fig.~\ref{fig:compactness}, early \casebh BSSs are connected by solid lines. Black holes formed by stars in the compactness peak (\cf Fig.~\ref{fig:compactness}) are highlighted by orange ellipses, and systems experiencing fallback are indicated by grey boxes.}
\end{figure*}

Despite the difference of a factor of 10 in metallicity, we find that the CO-core masses at the compactness peaks are similar ($M_\mathrm{CO}\approx7.5\,\msun$); the peak of single stars is shifted by $0.5\,\msun$ to lower $M_\mathrm{CO}$ compared to the $\zsun$ models and by $0.8\,\msun$ in \BSSs (Fig.~\ref{fig:compactness}). At lower metallicity with weaker winds, the convective helium-burning core grows relatively more in mass than at higher metallicity such that more $\alpha$ particles are mixed into it and more carbon is burnt into oxygen. With lower $X_\mathrm{C}$, the compactness peaks shift to smaller $M_\mathrm{CO}$ (Fig.~\ref{fig:compactness}).

At $Z=\zsun/10$, the second major increase in compactness of single stars is at almost the same $M_\mathrm{CO}$ of $\approx13\,\msun$ as in the $\zsun$ models. In \BSSs at $\zsun/10$, the transition is also at $M_\mathrm{CO}\approx13\,\msun$, but is smaller by $2\,\msun$ at $\zsun$. At the higher metallicity, wind mass loss during core helium burning in \BSSs with $M_\mathrm{CO}\gtrsim10\,\msun$ is so strong that the entire helium layer and parts of the CO-rich layer are lost, delaying the compactness-increase to higher $M_\mathrm{CO}$ and reducing the total stellar mass.

\section{\label{sec:compact-remnant-masses}Compact-remnant masses}

Applying our SN model, we find that stars in the compactness peak at $M_\mathrm{CO}\, {\approx}\, 7\text{--}8\,\msun$ and those with high compactness at $M_\mathrm{CO}\, {>}\, 13\text{--}15\,\msun$ produce BHs (Fig.~\ref{fig:compact-remnant-masses}). The same conclusion is reached by applying other explodability criteria based on neutrino-driven SNe: a compactness threshold of $0.44$ predicts the same explodability as in our analysis in ${>}\,95\%$ of cases and the two-parameter explodability criterion of \citet{2016ApJ...818..124E} agrees in ${>}\,92\%$ of cases (Appendix~\ref{app:stellar-models}). NS masses reach up to $2\,\msun$, where the lowest masses are from stars with the smallest compactness. In single stars, NS and BH formation occur at almost the same initial masses for both metallicities (at $M_\mathrm{ini}\,{\lesssim}\,35\,\msun$ and outside the compactness peak at $M_\mathrm{ini}\,{\approx}\,22\,\msun$; Fig.~\ref{fig:compact-remnant-masses}). In \BSSs, we find NS formation at $\zsun$ for initial masses of up to about $70\,\msun$ while this is up to about $40\,\msun$ at $\zsun/10$.

In general, at higher $Z$, winds are stronger such that the BH masses are smaller. Because \BSSs additionally lose their hydrogen-rich envelopes, their final and thus BH masses are smaller than those of single stars (Fig.~\ref{fig:compact-remnant-masses}).

The BH masses of \BSSs in the compactness peaks at $\zsun$ and $\zsun/10$ are $8.7\text{--}9.8\,\msun$, with the BHs at $\zsun/10$ being more massive by about $0.6\,\msun$ (Fig.~\ref{fig:compact-remnant-masses}). As shown above, the compactness peaks occur at a certain $M_\mathrm{CO}$, which is set by the size of the convective helium-burning core. This in turn is given by the total mass of the \BSS. Because the CO-core masses $M_\mathrm{CO}$ at the end of core helium of \BSSs in the compactness peak at $\zsun$ and $\zsun/10$ are similar, this implies that also the total helium-star mass and thus the BH mass must be similar, regardless of the exact mass loss history. Compactness-peak BHs from \BSSs thus have a characteristic mass across metallicity. The above argument no longer holds if mass loss during core helium burning reduces the helium core mass. The latter can happen in stars with $Z>\zsun$, and, beyond some $Z$, BH formation may no longer be possible \citep{2003ApJ...591..288H}.

The same argument holds for the BH masses of \BSSs of ${\approx}\,16\,\msun$ at the second major compactness-increase (Fig.~\ref{fig:compact-remnant-masses}). We have thus identified two characteristic BH masses from \BSSs of ${\approx}\,9\,\msun$ and ${\approx}\,16\,\msun$ that we call ``low-mass'' (\bhl) and ``high-mass'' (\bhh) BHs from hereon. 

\begin{figure*}
    \centering
    \includegraphics{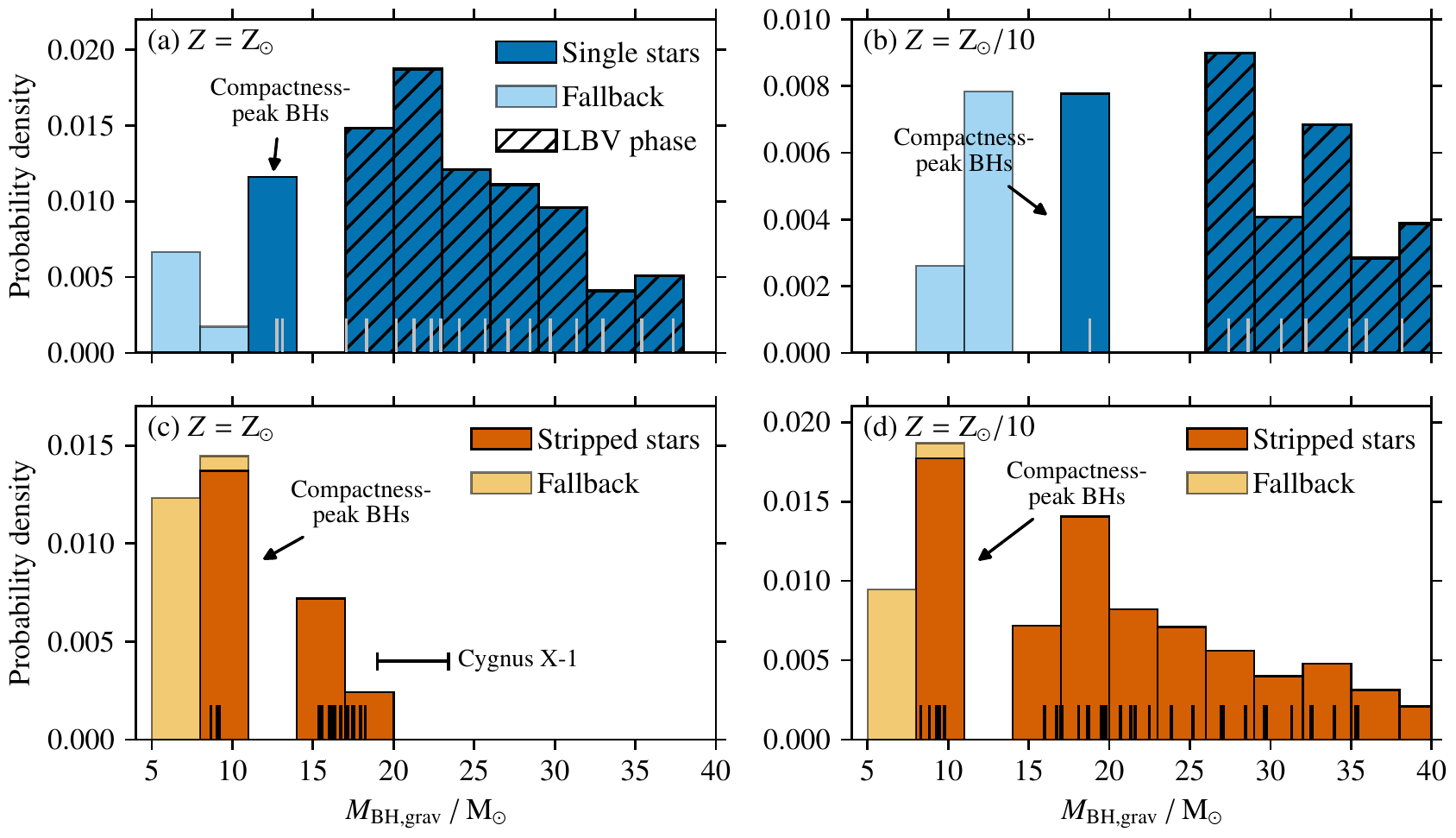}
    \caption{\label{fig:bh-mass-distribution}Black-hole mass distributions of the $\zsun$ (left panels) and $\zsun/10$ models (right panels), for single (top panels) and binary-stripped stars (bottom panels; \casea, B and~C combined). Black holes from the compactness peaks are indicated as well as systems evolving through LBV phases and thus likely undergoing enhanced mass loss. Grey and black ticks mark the individual masses of our models. A contribution from SN-fallback BHs is added assuming that 50\% a star's ejecta mass falls back onto the NS.}
\end{figure*}

In about 10--20\% of our $\zsun/10$ models beyond the compactness peak, the SN is not energetic enough to unbind the entire stellar envelope and SN fallback is expected. At $\zsun$, this fraction is 20--40\%. It remains unknown how much mass falls back and hence what is the resulting BH mass. Furthermore, some BH-forming stars may experience mass loss at core collapse because of neutrino losses leading to a weak SN-like transient \citep{1980Ap&SS..69..115N, 2013ApJ...769..109L}, and some BHs may accrete mass from their companion star during later binary-star evolution (\cf Fig.~\ref{fig:schematic-binary-evolution}). Such mechanisms are not accounted for in our models and can affect the BH masses.

\section{\label{sec:bh-mass-distribution}Black-hole mass distribution}

Assuming a Salpeter initial-mass function \citep{1955ApJ...121..161S} for single stars and primary stars in binaries, and that initial binary mass ratios and logarithmic orbital separations follow uniform distributions \citep{1924PTarO..25f...1O}, we compute the mass spectrum of BHs (Fig.~\ref{fig:bh-mass-distribution}). We mark single stars that spent more than $5\times10^4\,\mathrm{yr}$ in regions of the Hertzsprung--Russell diagram where S-Doradus luminous blue variables (LBVs) are found \citep{2004ApJ...615..475S, 2018MNRAS.478.3138D} because they likely experience unaccounted for mass loss. They may lose their entire envelope and behave like \BSSs.

Not considering SN fallback, the BH mass distributions at $\zsun$ and $\zsun/10$ are bimodal with a pronounced BH-mass gap and the lowest-mass BHs are from stars in the compactness peak (Fig.~\ref{fig:bh-mass-distribution}). In single stars, BHs are more massive than ${\approx}\,12\,\msun$ and the gap is at ${\approx}\,14\text{--}17\,\msun$ for $\zsun$ and at ${\approx}\,20\text{--}27\,\msun$ for $\zsun/10$. The masses of the compactness-peak BHs (${\approx}\,9\,\msun$) and the position of the BH mass gap at ${\approx}\,10\text{--}15\,\msun$ are similar in \BSSs at both metallicities. Compactness-peak BHs (\bhl) and BHs from the second major compactness increase (\bhh) can thus explain the inferred peaks at $9$ and $16\,\msun$ in the BH mass distribution of BBH mergers \citep{2021ApJ...913L..19T, 2021arXiv211103634T, 2022ApJ...928..155T}.

The BH mass spectrum of \BSSs at ${\gtrsim}20\,\msun$ depends on metallicity because of $Z$-dependent WR winds \citep[Fig.~\ref{fig:bh-mass-distribution};][]{2005A&A...442..587V, 2010ApJ...714.1217B, 2015MNRAS.451.4086S}. At $\zsun$, the maximum BH mass in our models barely reaches $20\,\msun$ while it is almost $45\,\msun$ at $\zsun/10$. This limits the maximum mass of BHs in X-ray binaries formed via binary-star evolution, and naturally explains why no BH with higher mass than in Cygnus X-1 has been found in the Milky Way ($21.2\pm2.2\,\msun$, \citealt{2021Sci...371.1046M}).

Assuming that 50\% of a star's envelope falls back, \BSS BHs in the range $5\text{--}8\,\msun$ and $6\text{--}9\,\msun$ form in our $\zsun$ and $\zsun/10$ models, respectively (Fig.~\ref{fig:bh-mass-distribution}). Fallback is thus required to explain the BH masses near the observationally-suggested NS-BH mass gap at $2\text{--}5\,\msun$ \citep{2010ApJ...725.1918O, 2011ApJ...741..103F, 2020A&A...636A..20W, 2021A&A...645A...5S}.

\section{\label{sec:chirp-mass-distribution}Chirp-mass distribution}

\begin{figure*}
    \centering
    \includegraphics{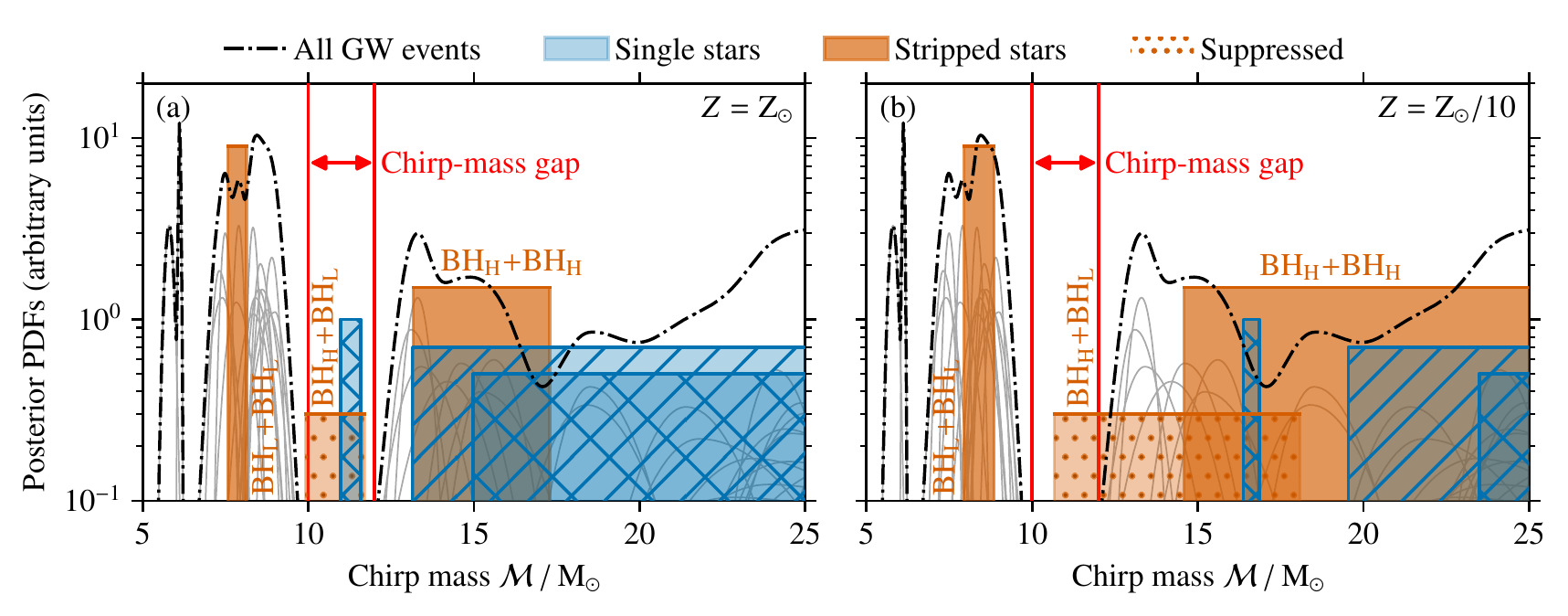}
    \caption{\label{fig:chirp-mass-distribution}Cumulative, non-normalized posterior probability distribution of the source chirp masses of 70 binary black-hole mergers with a false-alarm rate (FAR) of ${<}\,1\mathrm{yr}^{-1}$ \citep{2021arXiv211103634T}. The individual PDFs of all events are shown (thin grey lines) as well as the observed chirp-mass gap at $10\text{--}12\,\msun$. Chirp-mass ranges accessible by sampling BBH mergers from the BH mass distributions of single stars and \BSSs are indicated by colored boxes for $\zsun$ (panel a) and $\zsun/10$ (panel b; scaled arbitrarily in probability). For clarity, SN-fallback BHs broadening the \bhl{+}\bhl contribution are not shown. Chirp-mass ranges suppressed by isolated binary-star evolution are marked.}
\end{figure*}

The two newly-identified, characteristic BH masses of \BSSs of ${\approx}\,9\,\msun$ (\bhl) and ${\gtrsim}\,16\,\msun$ (\bhh) may lead to three features in the chirp-mass distribution of BBH mergers (\bhl{+}\bhl, \bhl{+}\bhh, and \bhh{+}\bhh). At $\zsun$, they are at $\mathcal{M}\approx7.5\text{--}8.1\,\msun$, $9.9\text{--}11.7\,\msun$ and $13.2\text{--}17.3\,\msun$, while they are at $\mathcal{M}\approx7.9\text{--}8.9\,\msun$, $10.7\text{--}18.1\,\msun$ and $14.6\text{--}42.2\,\msun$ at $\zsun/10$ (Fig.~\ref{fig:chirp-mass-distribution}). SN-fallback BHs will broaden the first contribution and extend it to the observed NS-BH mass gap.

Isolated binary-star evolution suppresses \bhl{+}\bhh mergers \citep[][Appendix~\ref{app:bse-suppression}]{2021A&A...645A...5S}. When forming the less-massive \bhl from stars in the compactness peak first, the initially less massive companion star cannot gain enough mass by mass transfer to result in a high-mass \bhh. A high-mass \bhh must form first, but then the parameter space for the companion to fall into the compactness peak and give rise to a low-mass \bhl is significantly smaller than forming another high-mass \bhh. Therefore, our models of \BSSs predict a dearth of BBH mergers at chirp masses of about $9\text{--}13\,\msun$.

\section{\label{sec:ns-bh-mergers}Neutron star--black hole and double neutron-star mergers}

We predict NS+\bhl mergers with $\mathcal{M}\approx2.7\text{--}3.7\,\msun$ and NS+\bhh mergers with $\mathcal{M}\approx 3.6\text{--}7.0 \,\msun$ from our \BSS models for a minimum NS mass of $1.3\,\msun$ and a maximum BH mass of $45\,\msun$. As explained above, forming the BHs in NS-BH mergers first is favored by binary-star evolution. Hence, NS-BH mergers with BH masses of $9\,\msun$ are predicted to occur more frequently than those with ${>}\,16\,\msun$ BHs because of the stellar initial mass function.

The NS-BH mergers GW200105 with component masses $8.9^{+1.2}_{-1.5}\,\msun$ and $1.9^{+0.3}_{-0.2}\,\msun$, and GW200115 with $5.7^{+1.8}_{-2.1}\,\msun$ and $1.5^{+0.7}_{-0.3}\,\msun$ \citep{2021ApJ...915L...5A} appear consistent with a compactness-peak and a SN-fallback BH, respectively. The high mass ratio event GW190814 of a $22.2\text{--}24.3\,\msun$ BH and a $2.50\text{--}2.67\,\msun$ compact object \citep{2020ApJ...896L..44A} might be an example of a high-mass \bhh merging with a very massive NS. Double NS mergers are possible with a total mass of up to $4\,\msun$ in our models, consistent with GW190425 \citep[total mass of $3.4^{+0.3}_{-0.1}\,\msun$,][]{2020ApJ...892L...3A}.

\section{\label{sec:discussion}Discussion and conclusion}

Predicting merger rates requires population-synthesis modelling including the cosmological star-formation history and thus the contributions of stellar populations of different $Z$. This is beyond the scope of this work. Nevertheless, because the \bhl and \bhh masses of \BSSs are so similar across a large range of metallicity, we predict peaks in the observed BH and chirp-mass distribution. This is in agreement with the location of the observed chirp-mass gap at $10\text{--}12\,\msun$ and also the peaks at $8$ and $14\,\msun$ in the chirp-mass distribution of 70 BBH mergers to date\footnote{Data taken from catalogues ``GWTC-1-confident'', ``GWTC-2.1-confident'' and ``GWTC-3-confident'', accessed via \url{https://www.gw-openscience.org} \citep{2021SoftX..1300658A}.} (Fig.~\ref{fig:chirp-mass-distribution}).

The peaks and gap in the BH mass distribution of single stars vary strongly with $Z$ and can thus not give rise to the peaks and gap in the observed chirp-mass distribution. Moreover, BBH mergers from single stars populate the observed chirp-mass gap and rather contribute at chirp masses ${\gtrsim}\,13\,\msun$ (Fig.~\ref{fig:chirp-mass-distribution}). Hence, the low chirp-mass regime in our models is dominated by BHs from \BSSs regardless of whether the BBH mergers are from the isolated-binary or dynamical-formation channel. This is not so surprising as \BSSs naturally have lower masses than single stars and thus also form less massive BHs. Furthermore, this interpretation is in line with models of dynamically-formed BBH mergers in dense stellar environments that have their dominant contribution at higher masses and underpredict the observed rate at ${\approx}\,10\,\msun$ \citep{2019PhRvD.100d3027R, 2019PhRvL.123r1101Y, 2023MNRAS.522..466A}. Hierarchical (repeated) BH mergers of \bhl and \bhh may help explain the observed peaks at $30$ and $57\,\msun$ in the BH mass distribution \citep{2021ApJ...913L..19T, 2022ApJ...928..155T, 2022arXiv220905766M}, and other mechanisms (\eg binary evolution) may add further features \citep{2022ApJ...940..184V} but leave our predicted characteristic BH masses unaltered.

The compactness landscape found in this work with a peak at $M_\mathrm{CO}\,{\approx}\,7\text{--}8\,\msun$ and high compactness at $M_\mathrm{CO}\,{\gtrsim}\,13\,\msun$ agrees with the results obtained by other groups using different stellar evolution codes and/or making different physics assumptions \citep[see \eg][]{2014ApJ...783...10S, 2020ApJ...890...43C, 2020MNRAS.499.2803P, 2021A&A...645A...5S, 2023ApJ...945...19T}. Moreover, \citet{2019ApJ...878...49W} and \citet{2023A&A...671A.134A} model helium stars at different metallicities as proxies of \BSSs and their results confirm that high compactness is found at characteristic $M_\mathrm{CO}$ for a large range of metallicities and that the final (\ie BH) masses of the helium stars at these $M_\mathrm{CO}$ are also similar, as in our work. \citet{2020ApJ...890...51E} study the collapse of the solar-metallicity helium stars of \citet{2019ApJ...878...49W} with their one-dimensional, neutrino-driven supernova code and confirm our results of a bimodal BH mass distribution with a large ${\approx}\,9\,\msun$ BH component from compactness-peak \BSSs and a gap at about $11\,\msun$ \citep[see also][]{2020ApJ...896...56W}.

The characteristic BH masses of \BSSs of $9$ and $16\,\msun$ and hence their bimodal BH mass spectrum depend on stellar, SN, and nuclear physics (see also Appendix~\ref{app:influence-bhs-physics}). Besides metallicity and wind strength (see above), the masses also depend, \eg, on convective core-boundary mixing and the $^{12}\mathrm{C}(\alpha,\gamma)^{16}\mathrm{O}$ nuclear reaction rate. While convective boundary mixing during core hydrogen burning mainly affects the number of BHs as it relates helium core masses to initial masses, the size of the convective helium-burning core directly affects the relation of CO-core to BH mass. Changing the convective core overshooting in our single-star models by a factor of 2 results in a $0.3\text{--}0.5\,\msun$ shift in $M_\mathrm{CO}$ of the compactness peak and thus a change in the BH masses of \BSSs by a similar amount. The $^{12}\mathrm{C}(\alpha,\gamma)^{16}\mathrm{O}$ reaction controls $X_\mathrm{C}$ and thus at which CO-core masses stars form BHs. Using the slower rate of \citet{2002ApJ...567..643K} leads to an increase of the compactness-peak $M_\mathrm{CO}$ by ${\approx}\,0.9\,\msun$. This reaction is also important for the pair-instability-SN BH mass gap \citep{2020ApJ...902L..36F}. The chirp-mass distribution will help constrain it and other essential physics determining the formation of stellar-mass BHs.

From GW detections of compact-object mergers, one can directly infer their luminosity distances and use them as ``standard sirens'' \citep{1986Natur.323..310S, 1993ApJ...411L...5C, 2005ApJ...629...15H}. Using universal features in the compact-object mass distribution, such as predicted here, one can then obtain the redshift to the sources and hence measure the cosmological redshift--luminosity distance relation. This allows for an independent determination of the Hubble constant \citep{2019ApJ...883L..42F, 2022PhRvL.129f1102E}. The characteristic BH masses identified here may thus help better understand the expansion history of the Universe.

\begin{acknowledgments}
FRNS and EL acknowledge support from the Klaus Tschira Foundation.
This work has received funding from the European Research Council (ERC) under the European Union’s Horizon 2020 research and innovation programme (Grant agreement No.\ 945806). This work is supported by the Deutsche Forschungsgemeinschaft (DFG, German Research Foundation) under Germany’s Excellence Strategy EXC 2181/1-390900948 (the Heidelberg STRUCTURES Excellence Cluster).
This research has made use of data or software obtained from the Gravitational Wave Open Science Center (gw-openscience.org), a service of LIGO Laboratory, the LIGO Scientific Collaboration, the Virgo Collaboration, and KAGRA. LIGO Laboratory and Advanced LIGO are funded by the United States National Science Foundation (NSF) as well as the Science and Technology Facilities Council (STFC) of the United Kingdom, the Max-Planck-Society (MPS), and the State of Niedersachsen/Germany for support of the construction of Advanced LIGO and construction and operation of the GEO600 detector. Additional support for Advanced LIGO was provided by the Australian Research Council. Virgo is funded, through the European Gravitational Observatory (EGO), by the French Centre National de Recherche Scientifique (CNRS), the Italian Istituto Nazionale di Fisica Nucleare (INFN) and the Dutch Nikhef, with contributions by institutions from Belgium, Germany, Greece, Hungary, Ireland, Japan, Monaco, Poland, Portugal, Spain. The construction and operation of KAGRA are funded by Ministry of Education, Culture, Sports, Science and Technology (MEXT), and Japan Society for the Promotion of Science (JSPS), National Research Foundation (NRF) and Ministry of Science and ICT (MSIT) in Korea, Academia Sinica (AS) and the Ministry of Science and Technology (MoST) in Taiwan.
\end{acknowledgments}

%

\vspace{5mm}


\software{NumPy \citep{oliphant2006numpy}, 
          SciPy \citep{2020NatMe..17..261V}, 
          Matplotlib \citep{hunter2007matplotlib},
          Jupyter Notebooks \citep{kluyver2016jupyternotebook}.
}



\clearpage
\appendix

\section{\label{app:stellar-models}Stellar models}

All models are non-rotating and the initial helium mass fraction of the newly computed models is $Y=0.24904$. This choice assumes that the helium abundance increases linearly over cosmic time, starting at $Y=0.24668$ set by Big Bang Nucleosynthesis \citep{2016A&A...594A..13P} and reaching the solar value of $Y_\odot=0.2703$ \citep{2009ARA&A..47..481A}. Convective boundary mixing during core hydrogen and core helium burning assumes step overshooting of $0.2$ pressure scale heights. The semi-convection efficiency is $0.1$, and the convective mixing-length parameter is $1.8$. Nuclear burning follows \mesa's \texttt{approx21\_cr60\_plus\_co56} reaction network and the nuclear reaction rates are taken from the JINA REACLIB database V2.2 \citep{2010ApJS..189..240C}, \ie the $^{12}\mathrm{C}(\alpha,\gamma)^{16}\mathrm{O}$ reaction rate is from the NACRE II compilation \citep{2013NuPhA.918...61X}. All models burn helium as red supergiants. Stellar wind mass-loss rates $\dot{M}$ are as in \mesa's ``Dutch'' wind scheme, but with a metallicity scaling of $\dot{M}\propto Z^{0.5}$ for stars with effective temperatures $T_\mathrm{eff}<10^4\,\mathrm{K}$ \citep{2011A&A...526A.156M} and following the model of \citep{2005A&A...442..587V} for winds of Wolf--Rayet (WR) stars. Models are computed until iron-core collapse. If stars do not directly collapse into BHs but emit neutrinos during the formation of a proto-NS, the neutrinos may carry away a significant fraction of the binding energy of a star such that some envelope mass is ejected \citep{1980Ap&SS..69..115N, 2013ApJ...769..109L}. This would systematically decrease our BH masses. Similarly, binary mass accretion after the formation of compact objects can in principle increase their mass. However, in the case of Eddington-limited accretion onto NSs and BHs, one may assume that the masses of NSs and BHs hardly change once they are formed.

To test the dependence of NS and BH formation in our models regarding the employed SN code, we apply two widely-used explodability criteria based on (i) the compactness $\xi_\mathrm{2.5}$ (Eq.~\ref{eq:xi}) and (ii) the two-parameter explodability criterion of \citet{2016ApJ...818..124E}. The latter uses the normalized mass inside a shell of specific entropy per nucleon of $s=4$,
\begin{equation}
    M_\mathrm{4} = \frac{m(s=4)}{\msun},
    \label{eq:M4}
\end{equation}
and the normalized mass derivative at this point in the star,
\begin{equation}
    \mu_\mathrm{4} = \frac{\mathrm{d}m/\msun}{\mathrm{d}r/1000\,\mathrm{km}}.
    \label{eq:mu4}
\end{equation}
Both criteria (as well as our SN code) rely on the neutrino-driven SN mechanism. For example, if stars explode via some engine-driven mechanism, the explodability criteria would be different (\eg they could be related to the rotational properties of a star at iron-core collapse). For a compactness threshold for BH formation of $\xi_\mathrm{2.5}>0.44$, we find the same explodability as in our analysis in ${>}\,95\%$ of cases, and the two-parameter explodability criterion of \citep{2016ApJ...818..124E}, $\mu_4 < k_1 M_4\mu_4 + k_2$ with constants $k_1=0.200$ and $k_2=0.072$, agrees in ${>}\,92\%$ of cases. Our results regarding successful SN explosions forming NSs and failed ones producing BHs thus seem robust against the exact explodability criteria.

Key stellar properties of the $\zsun$ and $\zsun/10$ models used throughout the paper are provided in Tables~\ref{tab:models_z} and~\ref{tab:models_z10}, respectively.

\onecolumngrid
\begin{ThreePartTable}
    \begin{longtable}[e]{@{\extracolsep{\fill}}cccccccccccccccc}
        \caption{\label{tab:models_z}Properties of the $\zsun$ models. Given are the initial mass $M_\mathrm{ini}$, model case, lifetime until core collapse $t_\mathrm{cc}$, final mass at core collapse $M_\mathrm{final}$, helium-core mass $M_\mathrm{He}$, CO-core mass $M_\mathrm{CO}$, iron-core mass $M_\mathrm{Fe}$, central mass fraction of carbon at the end of core helium burning $X_\mathrm{C}$, compactness $\xi_\mathrm{2.5}$, central specific entropy at core collapse $s_\mathrm{c}$, gravitational compact-remnant mass $M_\mathrm{rm,grav}$, radial mass derivative $\mu_4$ and mass $M_4$ at a specific entropy of $s=4$ (Eqs.~\ref{eq:mu4} and~\ref{eq:M4}), respectively, SN ejecta mass $M_\mathrm{ej}$, whether \casec mass transfer (MT) is possible and whether SN fallback is expected. The CO-core mass $M_\mathrm{CO}$ is given at the end of core helium burning and only changes marginally until core collapse. \casebe and~Bl mass transfer stands for ``early'' and ``late'' \caseb, \ie for mass transfer right after stars left the main sequence (radiative envelope) and before they ignite helium in their cores (convective envelope), respectively. Not all single and binary-stripped stars are evolved up to iron core collapse; in such cases, a blank row is reported in the table.}\\
        \toprule
        $M_\mathrm{ini}$ & Case & $t_\mathrm{cc}$ & $M_\mathrm{final}$ & $M_\mathrm{He}$ & $M_\mathrm{CO}$ & $M_\mathrm{Fe}$ & $X_\mathrm{C}$ & $\xi_\mathrm{2.5}$ & $s_\mathrm{c}$ & $M_\mathrm{rm,grav}$ & $\mu_\mathrm{4}$ & $M_\mathrm{4}$ & $M_\mathrm{ej}$ & \casec & Fall- \\
        ($\msun$) &  & (Myr) & ($\msun$) & ($\msun$) & ($\msun$) & ($\msun$) &  &  & ($N_\mathrm{A} k_\mathrm{B}$) & ($\msun$) &  &  & ($\msun$) & MT? & back? \\
        \midrule
        \endfirsthead
        \caption{continued.}\\
        \toprule
        $M_\mathrm{ini}$ & Case & $t_\mathrm{cc}$ & $M_\mathrm{final}$ & $M_\mathrm{He}$ & $M_\mathrm{CO}$ & $M_\mathrm{Fe}$ & $X_\mathrm{C}$ & $\xi_\mathrm{2.5}$ & $s_\mathrm{c}$ & $M_\mathrm{rm,grav}$ & $\mu_\mathrm{4}$ & $M_\mathrm{4}$ & $M_\mathrm{ej}$ & \casec & Fall- \\
        ($\msun$) &  & (Myr) & ($\msun$) & ($\msun$) & ($\msun$) & ($\msun$) &  &  & ($N_\mathrm{A} k_\mathrm{B}$) & ($\msun$) &  &  & ($\msun$) & MT? & back? \\
        \midrule
        \endhead
        \bottomrule
        \endfoot
         11.0 & Single & 22.4 & 10.0 & 3.3 & 1.9 & 1.51 & 0.33 & 0.01 & 0.78 & 1.30 & 0.02 & 1.53 & 8.6 & Yes & No \\
         11.0 & \casea & -- & -- & -- & -- & -- & -- & -- & -- & -- & -- & -- & -- & -- & -- \\
         11.0 & \casebe & -- & -- & -- & -- & -- & -- & -- & -- & -- & -- & -- & -- & -- & -- \\
         11.0 & \casebl & -- & -- & -- & -- & -- & -- & -- & -- & -- & -- & -- & -- & -- & -- \\
         11.0 & \casec & -- & -- & -- & -- & -- & -- & -- & -- & -- & -- & -- & -- & -- & -- \\
\midrule
         13.0 & Single & 17.0 & 11.3 & 4.3 & 2.6 & 1.57 & 0.30 & 0.08 & 0.86 & 1.42 & 0.04 & 1.65 & 9.7 & Yes & No \\
         13.0 & \casea & -- & -- & -- & -- & -- & -- & -- & -- & -- & -- & -- & -- & -- & -- \\
         13.0 & \casebe & -- & -- & -- & -- & -- & -- & -- & -- & -- & -- & -- & -- & -- & -- \\
         13.0 & \casebl & -- & -- & -- & -- & -- & -- & -- & -- & -- & -- & -- & -- & -- & -- \\
         13.0 & \casec & -- & -- & -- & -- & -- & -- & -- & -- & -- & -- & -- & -- & -- & -- \\
\midrule
         15.0 & Single & 13.8 & 12.2 & 5.2 & 3.4 & 1.65 & 0.29 & 0.11 & 0.86 & 1.40 & 0.04 & 1.63 & 10.6 & Yes & No \\
         15.0 & \casea & 14.7 & 3.4 & 3.4 & 1.9 & 1.49 & 0.36 & 0.02 & 0.80 & 1.30 & 0.03 & 1.52 & 1.9 & -- & No \\
         15.0 & \casebe & 14.2 & 3.3 & 3.3 & 1.9 & 1.53 & 0.35 & 0.02 & 0.80 & 1.32 & 0.03 & 1.50 & 1.8 & -- & No \\
         15.0 & \casebl & 14.2 & 3.4 & 3.4 & 1.9 & 1.50 & 0.36 & 0.02 & 0.82 & 1.34 & 0.03 & 1.53 & 1.9 & -- & No \\
         15.0 & \casec & 13.8 & 5.0 & 5.0 & 3.4 & 1.66 & 0.29 & 0.20 & 0.91 & 1.56 & 0.08 & 1.87 & 3.3 & -- & No \\
\midrule
         17.0 & Single & 11.7 & 13.0 & 6.3 & 4.2 & 1.77 & 0.27 & 0.20 & 0.95 & 1.54 & 0.07 & 1.82 & 11.3 & Yes & No \\
         17.0 & \casea & 12.4 & 3.9 & 3.9 & 2.4 & 1.57 & 0.35 & 0.05 & 0.81 & 1.33 & 0.03 & 1.67 & 2.5 & -- & No \\
         17.0 & \casebe & 11.9 & 4.0 & 4.0 & 2.4 & 1.62 & 0.35 & 0.06 & 0.80 & 1.32 & 0.03 & 1.69 & 2.5 & -- & No \\
         17.0 & \casebl & 11.9 & 4.0 & 4.0 & 2.5 & 1.55 & 0.35 & 0.06 & 0.84 & 1.35 & 0.03 & 1.70 & 2.5 & -- & No \\
         17.0 & \casec & 11.7 & 6.0 & 6.0 & 4.2 & 1.63 & 0.27 & 0.22 & 0.86 & 1.41 & 0.12 & 1.57 & 4.4 & -- & No \\
\midrule
         20.0 & Single & 9.6 & 13.0 & 7.9 & 5.6 & 1.80 & 0.25 & 0.24 & 0.97 & 1.58 & 0.08 & 1.86 & 11.2 & Yes & No \\
         20.0 & \casea & 10.0 & 4.9 & 4.9 & 3.3 & 1.61 & 0.33 & 0.16 & 0.90 & 1.46 & 0.07 & 1.72 & 3.3 & -- & No \\
         20.0 & \casebe & 9.7 & 5.0 & 5.0 & 3.4 & 1.58 & 0.33 & 0.13 & 0.88 & 1.41 & 0.05 & 1.65 & 3.4 & -- & No \\
         20.0 & \casebl & 9.7 & 5.0 & 5.0 & 3.4 & 1.53 & 0.33 & 0.07 & 0.83 & 1.36 & 0.03 & 1.59 & 3.5 & -- & No \\
         20.0 & \casec & 9.6 & 7.5 & 7.5 & 5.6 & 1.97 & 0.25 & 0.38 & 1.04 & 1.79 & 0.11 & 2.14 & 5.5 & -- & No \\
\midrule
         21.0 & Single & 9.1 & 12.9 & 8.5 & 6.1 & 1.60 & 0.25 & 0.18 & 0.91 & 1.44 & 0.07 & 1.69 & 11.2 & No & No \\
         21.0 & \casea & 9.5 & 5.2 & 5.2 & 3.6 & 1.55 & 0.32 & 0.09 & 0.84 & 1.37 & 0.04 & 1.60 & 3.7 & -- & No \\
         21.0 & \casebe & 9.2 & 5.3 & 5.3 & 3.7 & 1.63 & 0.32 & 0.14 & 0.87 & 1.41 & 0.06 & 1.67 & 3.7 & -- & No \\
         21.0 & \casebl & 9.2 & 5.3 & 5.3 & 3.8 & 1.63 & 0.32 & 0.17 & 0.91 & 1.45 & 0.06 & 1.73 & 3.7 & -- & No \\
         21.0 & \casec & 9.1 & 8.0 & 8.0 & 6.1 & 1.66 & 0.25 & 0.17 & 0.88 & 1.42 & 0.07 & 1.67 & 6.5 & -- & No \\
\midrule
         22.0 & Single & 8.7 & 13.1 & 9.0 & 6.6 & 2.04 & 0.24 & 0.57 & 1.16 & 13.1 & 0.15 & 2.36 & 0.0 & No & No \\
         22.0 & \casea & 9.0 & 5.5 & 5.5 & 3.9 & 1.66 & 0.32 & 0.17 & 0.90 & 1.53 & 0.06 & 1.88 & 3.8 & -- & No \\
         22.0 & \casebe & 8.8 & 5.6 & 5.6 & 4.0 & 1.57 & 0.32 & 0.09 & 0.81 & 1.35 & 0.03 & 1.55 & 4.1 & -- & No \\
         22.0 & \casebl & 8.8 & 5.7 & 5.7 & 4.1 & 1.73 & 0.31 & 0.21 & 0.94 & 1.52 & 0.07 & 1.80 & 4.0 & -- & No \\
         22.0 & \casec & 8.7 & 8.6 & 8.6 & 6.6 & 2.07 & 0.24 & 0.64 & 1.15 & 8.6 & 0.20 & 2.43 & 0.0 & -- & No \\
\midrule
         23.0 & Single & 8.3 & 12.7 & 9.6 & 7.1 & 2.07 & 0.23 & 0.59 & 1.11 & 12.7 & 0.19 & 2.37 & 0.0 & No & No \\
         23.0 & \casea & 8.6 & 5.8 & 5.8 & 4.3 & 1.65 & 0.31 & 0.17 & 0.91 & 1.51 & 0.06 & 1.87 & 4.1 & -- & No \\
         23.0 & \casebe & 8.4 & 5.9 & 5.9 & 4.4 & 1.66 & 0.31 & 0.20 & 0.93 & 1.48 & 0.08 & 1.75 & 4.3 & -- & No \\
         23.0 & \casebl & 8.4 & 6.0 & 6.0 & 4.5 & 1.57 & 0.31 & 0.13 & 0.87 & 1.43 & 0.05 & 1.69 & 4.4 & -- & No \\
         23.0 & \casec & 8.3 & 9.1 & 9.1 & 7.1 & 2.07 & 0.23 & 0.56 & 1.09 & 9.1 & 0.18 & 2.34 & 0.0 & -- & No \\
\midrule
         24.0 & Single & 7.9 & 13.2 & 10.1 & 7.5 & 1.89 & 0.23 & 0.40 & 1.05 & 1.77 & 0.12 & 2.07 & 11.2 & No & No \\
         24.0 & \casea & 8.2 & 6.1 & 6.1 & 4.6 & 1.59 & 0.31 & 0.16 & 0.90 & 1.45 & 0.06 & 1.71 & 4.5 & -- & No \\
         24.0 & \casebe & 8.0 & 6.2 & 6.2 & 4.7 & 1.59 & 0.30 & 0.20 & 0.89 & 1.41 & 0.08 & 1.65 & 4.7 & -- & No \\
         24.0 & \casebl & 8.0 & 6.3 & 6.3 & 4.8 & 1.64 & 0.30 & 0.21 & 0.93 & 1.49 & 0.08 & 1.76 & 4.7 & -- & No \\
         24.0 & \casec & 7.9 & 9.5 & 9.5 & 7.5 & 1.99 & 0.23 & 0.44 & 1.05 & 1.83 & 0.14 & 2.14 & 7.4 & -- & No \\
\midrule
         25.0 & Single & 7.6 & 13.3 & 10.6 & 8.0 & 1.65 & 0.22 & 0.19 & 0.94 & 1.53 & 0.06 & 1.80 & 11.6 & No & Yes \\
         25.0 & \casea & -- & -- & -- & -- & -- & -- & -- & -- & -- & -- & -- & -- & -- & -- \\
         25.0 & \casebe & -- & -- & -- & -- & -- & -- & -- & -- & -- & -- & -- & -- & -- & -- \\
         25.0 & \casebl & -- & -- & -- & -- & -- & -- & -- & -- & -- & -- & -- & -- & -- & -- \\
         25.0 & \casec & -- & -- & -- & -- & -- & -- & -- & -- & -- & -- & -- & -- & -- & -- \\
\midrule
         27.0 & Single & 7.1 & 13.9 & 11.8 & 9.1 & 1.85 & 0.21 & 0.20 & 0.95 & 1.57 & 0.08 & 1.85 & 12.1 & No & No \\
         27.0 & \casea & 7.3 & 7.0 & 7.0 & 5.6 & 1.80 & 0.29 & 0.28 & 1.02 & 1.66 & 0.08 & 1.96 & 5.2 & -- & No \\
         27.0 & \casebe & 7.2 & 7.2 & 7.2 & 5.8 & 1.64 & 0.28 & 0.20 & 0.93 & 1.50 & 0.08 & 1.75 & 5.5 & -- & No \\
         27.0 & \casebl & 7.1 & 7.3 & 7.3 & 5.9 & 1.75 & 0.28 & 0.21 & 0.97 & 1.64 & 0.05 & 2.09 & 5.4 & -- & No \\
         27.0 & \casec & 7.1 & 11.2 & 11.2 & 9.1 & 1.73 & 0.21 & 0.22 & 0.95 & 1.60 & 0.07 & 1.96 & 9.4 & -- & No \\
\midrule
         30.0 & Single & 6.5 & 14.6 & 13.4 & 10.7 & 1.66 & 0.20 & 0.24 & 0.94 & 1.51 & 0.08 & 1.79 & 12.9 & No & Yes \\
         30.0 & \casea & 6.7 & 7.8 & 7.8 & 6.5 & 1.68 & 0.28 & 0.18 & 0.94 & 1.51 & 0.06 & 1.79 & 6.1 & -- & No \\
         30.0 & \casebe & 6.5 & 8.2 & 8.2 & 7.0 & 1.73 & 0.27 & 0.17 & 0.90 & 1.47 & 0.07 & 1.73 & 6.5 & -- & No \\
         30.0 & \casebl & 6.5 & 8.2 & 8.2 & 7.0 & 1.76 & 0.27 & 0.20 & 0.92 & 1.50 & 0.08 & 1.74 & 6.5 & -- & No \\
         30.0 & \casec & 6.5 & 12.9 & 12.9 & 10.7 & 1.67 & 0.20 & 0.28 & 0.91 & 1.50 & 0.13 & 1.61 & 11.2 & -- & No \\
\midrule
         33.0 & Single & 6.0 & 16.0 & 16.0 & 12.4 & 1.89 & 0.19 & 0.44 & 1.09 & 1.83 & 0.13 & 2.14 & 13.8 & No & No \\
         33.0 & \casea & 6.1 & 8.7 & 8.7 & 7.6 & 2.20 & 0.26 & 0.66 & 1.15 & 8.7 & 0.19 & 2.51 & 0.0 & -- & No \\
         33.0 & \casebe & 6.0 & 9.0 & 9.0 & 8.1 & 2.10 & 0.26 & 0.60 & 1.11 & 9.0 & 0.19 & 2.37 & 0.0 & -- & No \\
         33.0 & \casebl & 6.0 & 9.1 & 9.1 & 8.2 & 2.12 & 0.26 & 0.60 & 1.11 & 9.1 & 0.21 & 2.36 & 0.0 & -- & No \\
         33.0 & \casec & 6.0 & 14.7 & 14.7 & 12.4 & 1.98 & 0.19 & 0.50 & 1.11 & 1.91 & 0.14 & 2.25 & 12.5 & -- & No \\
\midrule
         35.0 & Single & 5.7 & 17.1 & 17.1 & 13.5 & 2.14 & 0.18 & 0.68 & 1.17 & 17.1 & 0.25 & 2.42 & 0.0 & No & No \\
         35.0 & \casea & 5.9 & 9.2 & 9.2 & 8.3 & 2.09 & 0.26 & 0.60 & 1.11 & 9.2 & 0.21 & 2.31 & 0.0 & -- & No \\
         35.0 & \casebe & 5.8 & 9.6 & 9.6 & 8.8 & 1.83 & 0.25 & 0.34 & 1.01 & 1.66 & 0.11 & 1.96 & 7.7 & -- & No \\
         35.0 & \casebl & 5.8 & 9.5 & 9.5 & 8.8 & 1.66 & 0.25 & 0.20 & 0.94 & 1.52 & 0.06 & 1.80 & 7.8 & -- & Yes \\
         35.0 & \casec & 5.7 & 15.8 & 15.8 & 13.5 & 2.16 & 0.18 & 0.58 & 1.13 & 2.00 & 0.19 & 2.31 & 13.5 & -- & No \\
\midrule
         37.0 & Single & 5.5 & 18.3 & 18.3 & 14.5 & 2.23 & 0.17 & 0.70 & 1.18 & 18.3 & 0.32 & 2.34 & 0.0 & No & No \\
         37.0 & \casea & 5.6 & 9.7 & 9.7 & 9.0 & 1.69 & 0.25 & 0.21 & 0.95 & 1.54 & 0.06 & 1.83 & 8.0 & -- & Yes \\
         37.0 & \casebe & 5.5 & 9.6 & 9.6 & 9.6 & 1.71 & 0.24 & 0.16 & 0.91 & 1.46 & 0.05 & 1.73 & 8.0 & -- & Yes \\
         37.0 & \casebl & 5.5 & 9.6 & 9.6 & 9.6 & 1.68 & 0.24 & 0.20 & 0.93 & 1.51 & 0.06 & 1.77 & 7.9 & -- & Yes \\
         37.0 & \casec & 5.5 & 17.0 & 17.0 & 14.5 & 2.26 & 0.17 & 0.71 & 1.20 & 17.0 & 0.20 & 2.66 & 0.0 & -- & No \\
\midrule
         40.0 & Single & 5.2 & 20.2 & 20.2 & 16.2 & 2.23 & 0.17 & 0.74 & 1.25 & 20.2 & 0.17 & 2.58 & 0.0 & No & No \\
         40.0 & \casea & 5.3 & 10.2 & 10.2 & 9.9 & 1.61 & 0.24 & 0.23 & 0.87 & 1.40 & 0.10 & 1.60 & 8.7 & -- & No \\
         40.0 & \casebe & 5.2 & 10.2 & 10.2 & 10.3 & 1.71 & 0.24 & 0.25 & 0.97 & 1.57 & 0.08 & 1.84 & 8.4 & -- & Yes \\
         40.0 & \casebl & 5.2 & 10.2 & 10.2 & 10.2 & 1.76 & 0.24 & 0.28 & 0.91 & 1.54 & 0.15 & 1.60 & 8.4 & -- & No \\
         40.0 & \casec & 5.2 & 18.7 & 18.7 & 16.2 & 2.33 & 0.17 & 0.77 & 1.27 & 18.7 & 0.20 & 2.68 & 0.0 & -- & No \\
\midrule
         42.0 & Single & 5.0 & 21.3 & 21.3 & 17.3 & 2.63 & 0.16 & 0.78 & 1.27 & 21.3 & 0.22 & 2.83 & 0.0 & No & No \\
         42.0 & \casea & 5.2 & 10.4 & 10.4 & 10.6 & 1.64 & 0.24 & 0.23 & 0.89 & 1.41 & 0.09 & 1.64 & 8.9 & -- & No \\
         42.0 & \casebe & 5.1 & 10.6 & 10.6 & 10.7 & 1.66 & 0.23 & 0.20 & 0.92 & 1.56 & 0.06 & 1.98 & 8.9 & -- & No \\
         42.0 & \casebl & 5.1 & 10.7 & 10.7 & 10.8 & 1.72 & 0.23 & 0.27 & 0.97 & 1.63 & 0.08 & 2.00 & 8.9 & -- & No \\
         42.0 & \casec & 5.0 & 19.9 & 19.9 & 17.3 & 1.80 & 0.16 & 0.40 & 0.99 & 19.9 & 0.15 & 1.78 & 0.0 & -- & No \\
\midrule
         44.0 & Single & 4.9 & 22.3 & 22.3 & 18.4 & 2.57 & 0.16 & 0.77 & 1.30 & 22.3 & 0.48 & 1.84 & 0.0 & No & No \\
         44.0 & \casea & -- & -- & -- & -- & -- & -- & -- & -- & -- & -- & -- & -- & -- & -- \\
         44.0 & \casebe & -- & -- & -- & -- & -- & -- & -- & -- & -- & -- & -- & -- & -- & -- \\
         44.0 & \casebl & -- & -- & -- & -- & -- & -- & -- & -- & -- & -- & -- & -- & -- & -- \\
         44.0 & \casec & -- & -- & -- & -- & -- & -- & -- & -- & -- & -- & -- & -- & -- & -- \\
\midrule
         45.0 & Single & 4.8 & 22.9 & 22.9 & 18.9 & 2.67 & 0.15 & 0.80 & 1.37 & 22.9 & 0.51 & 1.96 & 0.0 & No & No \\
         45.0 & \casea & 4.9 & 10.9 & 10.9 & 11.0 & 1.78 & 0.23 & 0.28 & 0.98 & 1.65 & 0.08 & 2.04 & 9.0 & -- & No \\
         45.0 & \casebe & 4.9 & 11.4 & 11.4 & 11.4 & 1.72 & 0.23 & 0.24 & 0.97 & 1.65 & 0.07 & 2.03 & 9.5 & -- & No \\
         45.0 & \casebl & 4.9 & 11.5 & 11.5 & 11.5 & 1.59 & 0.23 & 0.18 & 0.84 & 1.37 & 0.10 & 1.57 & 9.9 & -- & No \\
         45.0 & \casec & 4.8 & 21.6 & 21.6 & 18.9 & 2.95 & 0.15 & 0.67 & 1.22 & 21.6 & 0.34 & 1.78 & 0.0 & -- & No \\
\midrule
         47.0 & Single & 4.7 & 24.1 & 24.1 & 20.0 & 2.54 & 0.15 & 0.73 & 1.31 & 24.1 & 0.45 & 1.72 & 0.0 & No & No \\
         47.0 & \casea & 4.8 & 11.4 & 11.4 & 11.4 & 1.72 & 0.23 & 0.24 & 0.95 & 1.60 & 0.14 & 1.74 & 9.5 & -- & No \\
         47.0 & \casebe & 4.7 & 11.8 & 11.8 & 11.8 & 1.60 & 0.23 & 0.18 & 0.89 & 1.43 & 0.07 & 1.67 & 10.2 & -- & Yes \\
         47.0 & \casebl & 4.7 & 11.8 & 11.8 & 11.9 & 1.60 & 0.22 & 0.18 & 0.90 & 1.45 & 0.07 & 1.69 & 10.2 & -- & Yes \\
         47.0 & \casec & 4.7 & 22.7 & 22.7 & 20.0 & 1.98 & 0.15 & 0.44 & 1.10 & 22.7 & 0.13 & 2.10 & 0.0 & -- & No \\
\midrule
         50.0 & Single & 4.5 & 25.7 & 25.7 & 21.6 & 1.91 & 0.14 & 0.44 & 1.11 & 25.7 & 0.13 & 2.08 & 0.0 & No & No \\
         50.0 & \casea & 4.6 & 12.0 & 12.0 & 12.0 & 1.75 & 0.22 & 0.25 & 1.00 & 1.64 & 0.10 & 1.93 & 10.1 & -- & No \\
         50.0 & \casebe & 4.6 & 12.4 & 12.4 & 12.5 & 1.78 & 0.22 & 0.22 & 0.95 & 1.52 & 0.08 & 1.80 & 10.7 & -- & Yes \\
         50.0 & \casebl & 4.6 & 12.5 & 12.5 & 12.5 & 1.72 & 0.22 & 0.26 & 0.99 & 1.60 & 0.10 & 1.88 & 10.6 & -- & No \\
         50.0 & \casec & 4.5 & 24.4 & 24.4 & 21.6 & 2.21 & 0.14 & 0.58 & 1.18 & 24.4 & 0.15 & 2.36 & 0.0 & -- & No \\
\midrule
         52.5 & Single & 4.4 & 27.1 & 27.1 & 23.0 & 2.02 & 0.14 & 0.55 & 1.19 & 27.1 & 0.15 & 2.30 & 0.0 & No & No \\
         52.5 & \casea & -- & -- & -- & -- & -- & -- & -- & -- & -- & -- & -- & -- & -- & -- \\
         52.5 & \casebe & -- & -- & -- & -- & -- & -- & -- & -- & -- & -- & -- & -- & -- & -- \\
         52.5 & \casebl & -- & -- & -- & -- & -- & -- & -- & -- & -- & -- & -- & -- & -- & -- \\
         52.5 & \casec & -- & -- & -- & -- & -- & -- & -- & -- & -- & -- & -- & -- & -- & -- \\
\midrule
         55.0 & Single & 4.3 & 28.5 & 28.5 & 24.3 & 2.15 & 0.14 & 0.57 & 1.18 & 28.5 & 0.15 & 2.32 & 0.0 & No & No \\
         55.0 & \casea & 4.4 & 13.0 & 13.0 & 13.0 & 1.74 & 0.21 & 0.27 & 1.00 & 1.60 & 0.09 & 1.89 & 11.1 & -- & Yes \\
         55.0 & \casebe & 4.3 & 13.3 & 13.3 & 13.3 & 1.79 & 0.21 & 0.35 & 1.05 & 1.68 & 0.11 & 2.00 & 11.3 & -- & No \\
         55.0 & \casebl & 4.3 & 13.2 & 13.2 & 13.3 & 1.90 & 0.21 & 0.30 & 0.99 & 1.62 & 0.10 & 1.90 & 11.4 & -- & No \\
         55.0 & \casec & 4.3 & 27.2 & 27.2 & 24.3 & 2.21 & 0.14 & 0.62 & 1.21 & 27.2 & 0.15 & 2.38 & 0.0 & -- & No \\
\midrule
         57.5 & Single & 4.2 & 29.7 & 29.7 & 25.7 & 2.21 & 0.13 & 0.63 & 1.21 & 29.7 & 0.15 & 2.40 & 0.0 & No & No \\
         57.5 & \casea & -- & -- & -- & -- & -- & -- & -- & -- & -- & -- & -- & -- & -- & -- \\
         57.5 & \casebe & -- & -- & -- & -- & -- & -- & -- & -- & -- & -- & -- & -- & -- & -- \\
         57.5 & \casebl & -- & -- & -- & -- & -- & -- & -- & -- & -- & -- & -- & -- & -- & -- \\
         57.5 & \casec & -- & -- & -- & -- & -- & -- & -- & -- & -- & -- & -- & -- & -- & -- \\
\midrule
         60.0 & Single & 4.1 & 31.4 & 31.4 & 26.9 & 2.17 & 0.13 & 0.58 & 1.20 & 31.4 & 0.16 & 2.29 & 0.0 & No & No \\
         60.0 & \casea & 4.2 & 13.3 & 13.3 & 13.3 & 1.83 & 0.21 & 0.31 & 1.01 & 1.65 & 0.11 & 1.94 & 11.4 & -- & No \\
         60.0 & \casebe & 4.1 & 13.9 & 13.9 & 14.0 & 1.87 & 0.21 & 0.28 & 0.98 & 1.60 & 0.09 & 1.87 & 12.1 & -- & Yes \\
         60.0 & \casebl & 4.1 & 13.9 & 13.9 & 14.0 & 1.84 & 0.21 & 0.33 & 1.02 & 1.65 & 0.10 & 1.96 & 12.0 & -- & No \\
         60.0 & \casec & 4.1 & 29.9 & 29.9 & 26.9 & 2.28 & 0.13 & 0.72 & 1.25 & 29.9 & 0.17 & 2.47 & 0.0 & -- & No \\
\midrule
         62.5 & Single & 4.0 & 33.0 & 33.0 & 28.3 & 2.32 & 0.13 & 0.67 & 1.24 & 33.0 & 0.17 & 2.42 & 0.0 & No & No \\
         62.5 & \casea & -- & -- & -- & -- & -- & -- & -- & -- & -- & -- & -- & -- & -- & -- \\
         62.5 & \casebe & -- & -- & -- & -- & -- & -- & -- & -- & -- & -- & -- & -- & -- & -- \\
         62.5 & \casebl & -- & -- & -- & -- & -- & -- & -- & -- & -- & -- & -- & -- & -- & -- \\
         62.5 & \casec & -- & -- & -- & -- & -- & -- & -- & -- & -- & -- & -- & -- & -- & -- \\
\midrule
         65.0 & Single & 3.9 & 35.4 & 35.4 & 29.6 & 3.04 & 0.12 & 0.87 & 1.49 & 35.4 & 0.29 & 2.41 & 0.0 & No & No \\
         65.0 & \casea & 4.0 & 14.3 & 14.3 & 14.4 & 1.91 & 0.20 & 0.34 & 1.05 & 1.70 & 0.10 & 2.02 & 12.4 & -- & No \\
         65.0 & \casebe & 4.0 & 14.7 & 14.7 & 14.8 & 1.69 & 0.20 & 0.28 & 0.94 & 1.51 & 0.10 & 1.74 & 13.0 & -- & No \\
         65.0 & \casebl & 4.0 & 14.8 & 14.8 & 14.8 & 1.89 & 0.20 & 0.41 & 1.09 & 1.76 & 0.12 & 2.10 & 12.7 & -- & No \\
         65.0 & \casec & 3.9 & 32.9 & 32.9 & 29.6 & 3.93 & 0.12 & 1.04 & 2.00 & 32.9 & 0.59 & 2.65 & 0.0 & -- & No \\
\midrule
         67.5 & Single & 3.9 & 37.3 & 37.3 & 31.0 & 2.47 & 0.12 & 0.82 & 1.36 & 37.3 & 0.24 & 2.59 & 0.0 & No & No \\
         67.5 & \casea & -- & -- & -- & -- & -- & -- & -- & -- & -- & -- & -- & -- & -- & -- \\
         67.5 & \casebe & -- & -- & -- & -- & -- & -- & -- & -- & -- & -- & -- & -- & -- & -- \\
         67.5 & \casebl & -- & -- & -- & -- & -- & -- & -- & -- & -- & -- & -- & -- & -- & -- \\
         67.5 & \casec & -- & -- & -- & -- & -- & -- & -- & -- & -- & -- & -- & -- & -- & -- \\
\midrule
         70.0 & Single & 3.8 & 41.4 & 41.4 & 32.4 & 2.42 & 0.12 & 0.82 & 1.33 & 41.4 & 0.19 & 2.78 & 0.0 & No & No \\
         70.0 & \casea & 3.9 & 14.9 & 14.9 & 15.0 & 1.87 & 0.20 & 0.38 & 1.07 & 1.72 & 0.11 & 2.06 & 12.9 & -- & No \\
         70.0 & \casebe & 3.8 & 15.4 & 15.4 & 15.4 & 2.06 & 0.20 & 0.66 & 1.16 & 15.4 & 0.21 & 2.46 & 0.0 & -- & No \\
         70.0 & \casebl & 3.8 & 15.5 & 15.5 & 15.6 & 2.13 & 0.19 & 0.66 & 1.15 & 15.5 & 0.26 & 2.39 & 0.0 & -- & No \\
         70.0 & \casec & 3.8 & 35.8 & 35.8 & 32.4 & 2.81 & 0.12 & 0.82 & 1.36 & 35.8 & 0.23 & 2.93 & 0.0 & -- & No \\
\midrule
         72.5 & Single & 3.7 & 45.7 & 45.7 & 33.7 & 2.60 & 0.11 & 0.85 & 1.43 & 45.7 & 0.35 & 2.85 & 0.0 & No & No \\
         72.5 & \casea & -- & -- & -- & -- & -- & -- & -- & -- & -- & -- & -- & -- & -- & -- \\
         72.5 & \casebe & -- & -- & -- & -- & -- & -- & -- & -- & -- & -- & -- & -- & -- & -- \\
         72.5 & \casebl & -- & -- & -- & -- & -- & -- & -- & -- & -- & -- & -- & -- & -- & -- \\
         72.5 & \casec & -- & -- & -- & -- & -- & -- & -- & -- & -- & -- & -- & -- & -- & -- \\
\midrule
         75.0 & Single & 3.7 & 47.0 & 47.0 & 35.0 & 2.56 & 0.11 & 0.83 & 1.39 & 47.0 & 0.31 & 2.77 & 0.0 & No & No \\
         75.0 & \casea & 3.8 & 15.3 & 15.3 & 15.4 & 2.03 & 0.20 & 0.59 & 1.15 & 15.3 & 0.16 & 2.39 & 0.0 & -- & No \\
         75.0 & \casebe & 3.7 & 16.2 & 16.2 & 16.3 & 2.20 & 0.19 & 0.68 & 1.17 & 16.2 & 0.29 & 2.38 & 0.0 & -- & No \\
         75.0 & \casebl & 3.7 & 16.2 & 16.2 & 16.3 & 2.18 & 0.19 & 0.69 & 1.17 & 16.2 & 0.19 & 2.57 & 0.0 & -- & No \\
         75.0 & \casec & -- & -- & -- & -- & -- & -- & -- & -- & -- & -- & -- & -- & -- & -- \\
\midrule
         80.0 & Single & -- & -- & -- & -- & -- & -- & -- & -- & -- & -- & -- & -- & -- & -- \\
         80.0 & \casea & 3.7 & 16.0 & 16.0 & 16.1 & 2.13 & 0.19 & 0.69 & 1.19 & 16.0 & 0.21 & 2.51 & 0.0 & -- & No \\
         80.0 & \casebe & 3.6 & 16.7 & 16.7 & 16.8 & 2.21 & 0.19 & 0.69 & 1.19 & 16.7 & 0.32 & 2.31 & 0.0 & -- & No \\
         80.0 & \casebl & 3.6 & 17.1 & 17.1 & 17.2 & 2.24 & 0.18 & 0.72 & 1.20 & 17.1 & 0.33 & 2.38 & 0.0 & -- & No \\
         80.0 & \casec & -- & -- & -- & -- & -- & -- & -- & -- & -- & -- & -- & -- & -- & -- \\
\midrule
         85.0 & Single & -- & -- & -- & -- & -- & -- & -- & -- & -- & -- & -- & -- & -- & -- \\
         85.0 & \casea & 3.6 & 16.1 & 16.1 & 16.2 & 2.13 & 0.19 & 0.65 & 1.15 & 16.1 & 0.29 & 2.24 & 0.0 & -- & No \\
         85.0 & \casebe & 3.5 & 17.4 & 17.4 & 17.5 & 2.30 & 0.18 & 0.72 & 1.21 & 17.4 & 0.34 & 2.39 & 0.0 & -- & No \\
         85.0 & \casebl & 3.5 & 17.6 & 17.6 & 17.6 & 2.30 & 0.18 & 0.73 & 1.23 & 17.6 & 0.35 & 2.40 & 0.0 & -- & No \\
         85.0 & \casec & -- & -- & -- & -- & -- & -- & -- & -- & -- & -- & -- & -- & -- & -- \\
\midrule
         90.0 & Single & -- & -- & -- & -- & -- & -- & -- & -- & -- & -- & -- & -- & -- & -- \\
         90.0 & \casea & 3.5 & 16.4 & 16.4 & 16.4 & 2.17 & 0.19 & 0.67 & 1.16 & 16.4 & 0.30 & 2.27 & 0.0 & -- & No \\
         90.0 & \casebe & 3.4 & 17.9 & 17.9 & 18.0 & 2.40 & 0.18 & 0.76 & 1.24 & 17.9 & 0.35 & 2.55 & 0.0 & -- & No \\
         90.0 & \casebl & 3.4 & 18.2 & 18.2 & 18.3 & 2.38 & 0.18 & 0.74 & 1.24 & 18.2 & 0.36 & 2.41 & 0.0 & -- & No \\
         90.0 & \casec & -- & -- & -- & -- & -- & -- & -- & -- & -- & -- & -- & -- & -- & -- \\
    \end{longtable}
\end{ThreePartTable}
\begin{ThreePartTable}
    \begin{longtable}[e]{@{\extracolsep{\fill}}cccccccccccccccc}
        \caption{\label{tab:models_z10}Same as Table~\ref{tab:models_z} but for the $\zsun/10$ models.}\\
        \toprule
        $M_\mathrm{ini}$ & Case & $t_\mathrm{cc}$ & $M_\mathrm{final}$ & $M_\mathrm{He}$ & $M_\mathrm{CO}$ & $M_\mathrm{Fe}$ & $X_\mathrm{C}$ & $\xi_\mathrm{2.5}$ & $s_\mathrm{c}$ & $M_\mathrm{rm,grav}$ & $\mu_\mathrm{4}$ & $M_\mathrm{4}$ & $M_\mathrm{ej}$ & \casec & Fall- \\
        ($\msun$) &  & (Myr) & ($\msun$) & ($\msun$) & ($\msun$) & ($\msun$) &  &  & ($N_\mathrm{A} k_\mathrm{B}$) & ($\msun$) &  &  & ($\msun$) & MT? & back? \\
        \midrule
        \endfirsthead
        \caption{continued.}\\
        \toprule
        $M_\mathrm{ini}$ & Case & $t_\mathrm{cc}$ & $M_\mathrm{final}$ & $M_\mathrm{He}$ & $M_\mathrm{CO}$ & $M_\mathrm{Fe}$ & $X_\mathrm{C}$ & $\xi_\mathrm{2.5}$ & $s_\mathrm{c}$ & $M_\mathrm{rm,grav}$ & $\mu_\mathrm{4}$ & $M_\mathrm{4}$ & $M_\mathrm{ej}$ & \casec & Fall- \\
        ($\msun$) &  & (Myr) & ($\msun$) & ($\msun$) & ($\msun$) & ($\msun$) &  &  & ($N_\mathrm{A} k_\mathrm{B}$) & ($\msun$) &  &  & ($\msun$) & MT? & back? \\
        \midrule
        \endhead
        \bottomrule
        \endfoot
         11.0 & Single & 22.3 & 10.8 & 3.8 & 2.2 & 1.55 & 0.29 & 0.04 & 0.84 & 1.38 & 0.04 & 1.63 & 9.3 & Yes & No \\
         11.0 & \casea & -- & -- & -- & -- & -- & -- & -- & -- & -- & -- & -- & -- & -- & -- \\
         11.0 & \casebe & -- & -- & -- & -- & -- & -- & -- & -- & -- & -- & -- & -- & -- & -- \\
         11.0 & \casebl & -- & -- & -- & -- & -- & -- & -- & -- & -- & -- & -- & -- & -- & -- \\
         11.0 & \casec & -- & -- & -- & -- & -- & -- & -- & -- & -- & -- & -- & -- & -- & -- \\
\midrule
         13.0 & Single & 17.3 & 12.7 & 4.7 & 2.9 & 1.59 & 0.26 & 0.12 & 0.89 & 1.44 & 0.05 & 1.70 & 11.1 & Yes & No \\
         13.0 & \casea & -- & -- & -- & -- & -- & -- & -- & -- & -- & -- & -- & -- & -- & -- \\
         13.0 & \casebe & -- & -- & -- & -- & -- & -- & -- & -- & -- & -- & -- & -- & -- & -- \\
         13.0 & \casebl & -- & -- & -- & -- & -- & -- & -- & -- & -- & -- & -- & -- & -- & -- \\
         13.0 & \casec & -- & -- & -- & -- & -- & -- & -- & -- & -- & -- & -- & -- & -- & -- \\
\midrule
         15.0 & Single & 14.2 & 14.2 & 5.5 & 3.6 & 1.59 & 0.28 & 0.14 & 0.90 & 1.45 & 0.05 & 1.73 & 12.6 & Yes & No \\
         15.0 & \casea & 14.9 & 4.0 & 4.0 & 2.4 & 1.53 & 0.32 & 0.06 & 0.83 & 1.36 & 0.04 & 1.57 & 2.5 & -- & No \\
         15.0 & \casebe & 14.5 & 3.8 & 3.8 & 2.3 & 1.51 & 0.33 & 0.04 & 0.83 & 1.35 & 0.04 & 1.58 & 2.3 & -- & No \\
         15.0 & \casebl & 14.5 & 3.9 & 3.9 & 2.3 & 1.51 & 0.33 & 0.05 & 0.82 & 1.34 & 0.04 & 1.55 & 2.4 & -- & No \\
         15.0 & \casec & 14.2 & 5.3 & 5.3 & 3.6 & 1.62 & 0.28 & 0.14 & 0.90 & 1.50 & 0.05 & 1.83 & 3.6 & -- & No \\
\midrule
         17.0 & Single & 12.1 & 15.9 & 6.6 & 4.5 & 1.75 & 0.26 & 0.25 & 0.98 & 1.65 & 0.09 & 1.91 & 14.0 & Yes & No \\
         17.0 & \casea & 12.6 & 5.0 & 5.0 & 3.2 & 1.55 & 0.30 & 0.15 & 0.81 & 1.33 & 0.09 & 1.52 & 3.5 & -- & No \\
         17.0 & \casebe & 12.3 & 4.7 & 4.7 & 3.0 & 1.60 & 0.31 & 0.14 & 0.89 & 1.45 & 0.06 & 1.69 & 3.0 & -- & No \\
         17.0 & \casebl & 12.3 & 4.7 & 4.7 & 3.0 & 1.58 & 0.31 & 0.13 & 0.89 & 1.42 & 0.06 & 1.67 & 3.1 & -- & No \\
         17.0 & \casec & 12.1 & 6.2 & 6.2 & 4.5 & 1.79 & 0.26 & 0.20 & 0.95 & 1.56 & 0.07 & 1.82 & 4.5 & -- & No \\
\midrule
         20.0 & Single & 10.0 & 18.1 & 8.3 & 5.9 & 1.72 & 0.24 & 0.23 & 0.95 & 1.61 & 0.13 & 1.79 & 16.3 & Yes & No \\
         20.0 & \casea & 10.3 & 6.3 & 6.3 & 4.3 & 1.62 & 0.29 & 0.17 & 0.93 & 1.49 & 0.06 & 1.76 & 4.6 & -- & No \\
         20.0 & \casebe & 10.1 & 6.0 & 6.0 & 4.1 & 1.60 & 0.29 & 0.15 & 0.91 & 1.47 & 0.05 & 1.72 & 4.3 & -- & No \\
         20.0 & \casebl & 10.1 & 6.1 & 6.1 & 4.2 & 1.68 & 0.29 & 0.15 & 0.91 & 1.46 & 0.06 & 1.73 & 4.4 & -- & No \\
         20.0 & \casec & 10.0 & 7.8 & 7.8 & 5.9 & 1.60 & 0.24 & 0.18 & 0.86 & 1.38 & 0.08 & 1.59 & 6.2 & -- & No \\
\midrule
         21.0 & Single & 9.5 & 18.8 & 8.9 & 6.3 & 2.11 & 0.23 & 0.68 & 1.17 & 18.8 & 0.27 & 2.37 & 0.0 & Yes & No \\
         21.0 & \casea & 9.8 & 6.8 & 6.8 & 4.8 & 1.79 & 0.28 & 0.32 & 1.01 & 1.66 & 0.10 & 1.96 & 4.9 & -- & No \\
         21.0 & \casebe & 9.6 & 6.4 & 6.4 & 4.5 & 1.71 & 0.28 & 0.23 & 0.95 & 1.55 & 0.08 & 1.84 & 4.7 & -- & No \\
         21.0 & \casebl & 9.6 & 6.5 & 6.5 & 4.6 & 1.77 & 0.28 & 0.33 & 1.02 & 1.68 & 0.11 & 1.98 & 4.6 & -- & No \\
         21.0 & \casec & 9.5 & 8.3 & 8.3 & 6.3 & 2.06 & 0.23 & 0.62 & 1.16 & 8.3 & 0.17 & 2.38 & 0.0 & -- & No \\
\midrule
         22.0 & Single & 9.0 & 19.5 & 9.5 & 6.8 & 1.89 & 0.23 & 0.42 & 1.08 & 1.81 & 0.12 & 2.12 & 17.4 & Yes & No \\
         22.0 & \casea & 9.3 & 7.3 & 7.3 & 5.2 & 1.78 & 0.27 & 0.35 & 1.04 & 1.67 & 0.12 & 1.97 & 5.4 & -- & No \\
         22.0 & \casebe & 9.1 & 6.9 & 6.9 & 4.9 & 1.85 & 0.28 & 0.34 & 1.02 & 1.69 & 0.11 & 2.02 & 5.0 & -- & No \\
         22.0 & \casebl & 9.1 & 7.0 & 7.0 & 5.0 & 1.73 & 0.28 & 0.28 & 0.99 & 1.59 & 0.10 & 1.87 & 5.2 & -- & No \\
         22.0 & \casec & 9.0 & 8.8 & 8.8 & 6.8 & 2.07 & 0.23 & 0.61 & 1.12 & 8.8 & 0.25 & 2.26 & 0.0 & -- & No \\
\midrule
         23.0 & Single & 8.6 & 20.1 & 10.0 & 7.3 & 1.85 & 0.22 & 0.38 & 1.04 & 1.74 & 0.12 & 2.04 & 18.1 & Yes & No \\
         23.0 & \casea & 8.9 & 7.7 & 7.7 & 5.6 & 1.86 & 0.27 & 0.26 & 0.97 & 1.59 & 0.09 & 1.86 & 5.9 & -- & No \\
         23.0 & \casebe & 8.7 & 7.4 & 7.4 & 5.4 & 1.73 & 0.27 & 0.27 & 0.99 & 1.62 & 0.10 & 1.91 & 5.5 & -- & No \\
         23.0 & \casebl & 8.7 & 7.5 & 7.5 & 5.4 & 1.68 & 0.27 & 0.24 & 0.96 & 1.54 & 0.09 & 1.81 & 5.7 & -- & No \\
         23.0 & \casec & 8.6 & 9.4 & 9.4 & 7.3 & 1.99 & 0.22 & 0.43 & 1.07 & 1.86 & 0.13 & 2.17 & 7.2 & -- & No \\
\midrule
         24.0 & Single & 8.3 & 20.7 & 10.6 & 7.9 & 1.78 & 0.22 & 0.18 & 0.91 & 1.52 & 0.05 & 1.80 & 19.0 & Yes & Yes \\
         24.0 & \casea & 8.5 & 8.3 & 8.3 & 6.1 & 1.60 & 0.26 & 0.12 & 0.85 & 1.37 & 0.06 & 1.62 & 6.7 & -- & No \\
         24.0 & \casebe & 8.3 & 7.8 & 7.8 & 5.8 & 1.75 & 0.27 & 0.17 & 0.91 & 1.49 & 0.07 & 1.75 & 6.1 & -- & No \\
         24.0 & \casebl & 8.3 & 8.0 & 8.0 & 5.9 & 1.62 & 0.26 & 0.17 & 0.87 & 1.41 & 0.06 & 1.63 & 6.4 & -- & No \\
         24.0 & \casec & 8.3 & 9.9 & 9.9 & 7.9 & 1.63 & 0.22 & 0.17 & 0.93 & 1.50 & 0.06 & 1.75 & 8.2 & -- & Yes \\
\midrule
         25.0 & Single & 7.9 & 21.3 & 11.1 & 8.4 & 1.66 & 0.21 & 0.23 & 0.93 & 1.50 & 0.08 & 1.77 & 19.6 & No & Yes \\
         25.0 & \casea & 8.1 & 8.7 & 8.7 & 6.6 & 1.79 & 0.26 & 0.33 & 1.04 & 1.68 & 0.11 & 1.98 & 6.8 & -- & No \\
         25.0 & \casebe & 8.0 & 8.3 & 8.3 & 6.3 & 1.72 & 0.26 & 0.16 & 0.90 & 1.47 & 0.07 & 1.72 & 6.7 & -- & No \\
         25.0 & \casebl & 8.0 & 8.5 & 8.5 & 6.4 & 1.59 & 0.26 & 0.18 & 0.90 & 1.45 & 0.07 & 1.69 & 6.9 & -- & No \\
         25.0 & \casec & 7.9 & 10.5 & 10.5 & 8.4 & 1.75 & 0.21 & 0.21 & 0.92 & 1.50 & 0.07 & 1.77 & 8.8 & -- & Yes \\
\midrule
         26.0 & Single & 7.7 & 21.9 & 11.6 & 8.9 & 1.82 & 0.21 & 0.20 & 0.94 & 1.55 & 0.08 & 1.83 & 20.1 & No & No \\
         26.0 & \casea & 7.8 & 9.3 & 9.3 & 7.1 & 2.14 & 0.25 & 0.66 & 1.17 & 9.3 & 0.18 & 2.49 & 0.0 & -- & No \\
         26.0 & \casebe & 7.7 & 8.8 & 8.8 & 6.7 & 2.08 & 0.25 & 0.49 & 1.11 & 1.94 & 0.14 & 2.23 & 6.6 & -- & No \\
         26.0 & \casebl & 7.7 & 8.9 & 8.9 & 6.8 & 1.96 & 0.25 & 0.50 & 1.11 & 1.92 & 0.14 & 2.23 & 6.7 & -- & No \\
         26.0 & \casec & 7.7 & 11.0 & 11.0 & 8.9 & 1.80 & 0.21 & 0.19 & 0.94 & 1.52 & 0.08 & 1.80 & 9.3 & -- & No \\
\midrule
         27.0 & Single & 7.4 & 22.4 & 12.2 & 9.4 & 1.89 & 0.20 & 0.25 & 0.92 & 1.59 & 0.16 & 1.58 & 20.6 & No & No \\
         27.0 & \casea & 7.6 & 9.8 & 9.8 & 7.5 & 2.08 & 0.24 & 0.55 & 1.10 & 9.8 & 0.17 & 2.31 & 0.0 & -- & No \\
         27.0 & \casebe & 7.5 & 9.3 & 9.3 & 7.1 & 2.09 & 0.25 & 0.65 & 1.14 & 9.3 & 0.22 & 2.42 & 0.0 & -- & No \\
         27.0 & \casebl & 7.5 & 9.5 & 9.5 & 7.3 & 2.10 & 0.25 & 0.63 & 1.13 & 9.5 & 0.23 & 2.36 & 0.0 & -- & No \\
         27.0 & \casec & 7.4 & 11.6 & 11.6 & 9.4 & 1.79 & 0.20 & 0.21 & 0.93 & 1.53 & 0.08 & 1.80 & 9.9 & -- & No \\
\midrule
         30.0 & Single & 6.7 & 24.1 & 14.1 & 11.0 & 1.67 & 0.19 & 0.28 & 0.91 & 1.48 & 0.12 & 1.66 & 22.5 & No & No \\
         30.0 & \casea & 6.9 & 11.3 & 11.3 & 9.0 & 1.72 & 0.23 & 0.27 & 0.97 & 1.63 & 0.08 & 2.00 & 9.5 & -- & No \\
         30.0 & \casebe & 6.8 & 10.8 & 10.8 & 8.6 & 1.73 & 0.23 & 0.24 & 0.89 & 1.48 & 0.14 & 1.57 & 9.1 & -- & No \\
         30.0 & \casebl & 6.8 & 11.0 & 11.0 & 8.7 & 1.68 & 0.23 & 0.15 & 0.88 & 1.44 & 0.06 & 1.69 & 9.3 & -- & No \\
         30.0 & \casec & 6.7 & 13.3 & 13.3 & 11.0 & 1.75 & 0.19 & 0.30 & 1.00 & 1.61 & 0.10 & 1.90 & 11.5 & -- & No \\
\midrule
         33.0 & Single & 6.2 & 26.2 & 15.9 & 12.6 & 1.93 & 0.18 & 0.48 & 1.13 & 1.89 & 0.13 & 2.21 & 24.0 & No & No \\
         33.0 & \casea & 6.3 & 13.0 & 13.0 & 10.5 & 1.67 & 0.22 & 0.22 & 0.94 & 1.52 & 0.09 & 1.79 & 11.3 & -- & No \\
         33.0 & \casebe & 6.3 & 12.4 & 12.4 & 10.1 & 2.23 & 0.22 & 0.23 & 0.92 & 1.54 & 0.15 & 1.58 & 10.6 & -- & No \\
         33.0 & \casebl & 6.3 & 12.5 & 12.5 & 10.2 & 1.82 & 0.22 & 0.22 & 0.97 & 1.60 & 0.09 & 1.89 & 10.7 & -- & No \\
         33.0 & \casec & 6.2 & 15.0 & 15.0 & 12.6 & 2.08 & 0.18 & 0.62 & 1.14 & 15.0 & 0.18 & 2.44 & 0.0 & -- & No \\
\midrule
         35.0 & Single & 6.0 & 27.4 & 16.8 & 13.7 & 2.14 & 0.18 & 0.70 & 1.18 & 27.4 & 0.28 & 2.42 & 0.0 & No & No \\
         35.0 & \casea & 6.1 & 14.0 & 14.0 & 11.6 & 1.77 & 0.21 & 0.31 & 1.03 & 1.63 & 0.10 & 1.95 & 12.2 & -- & No \\
         35.0 & \casebe & 6.0 & 13.4 & 13.4 & 11.1 & 1.65 & 0.21 & 0.23 & 0.95 & 1.51 & 0.08 & 1.77 & 11.7 & -- & Yes \\
         35.0 & \casebl & 6.0 & 13.5 & 13.5 & 11.1 & 1.73 & 0.21 & 0.28 & 1.01 & 1.62 & 0.09 & 1.91 & 11.6 & -- & No \\
         35.0 & \casec & 6.0 & 16.2 & 16.2 & 13.7 & 2.13 & 0.18 & 0.68 & 1.19 & 16.2 & 0.18 & 2.51 & 0.0 & -- & No \\
\midrule
         37.0 & Single & 5.7 & 28.6 & 21.0 & 14.2 & 2.08 & 0.16 & 0.64 & 1.21 & 28.6 & 0.17 & 2.38 & 0.0 & No & No \\
         37.0 & \casea & 5.8 & 15.1 & 15.1 & 12.5 & 1.83 & 0.20 & 0.35 & 1.04 & 1.69 & 0.11 & 1.99 & 13.1 & -- & No \\
         37.0 & \casebe & 5.8 & 14.4 & 14.4 & 12.0 & 1.88 & 0.21 & 0.35 & 1.04 & 1.72 & 0.11 & 2.02 & 12.4 & -- & No \\
         37.0 & \casebl & 5.7 & 14.5 & 14.5 & 12.1 & 1.73 & 0.21 & 0.29 & 1.00 & 1.61 & 0.10 & 1.91 & 12.7 & -- & No \\
         37.0 & \casec & 5.7 & 16.9 & 16.9 & 14.2 & 2.37 & 0.16 & 0.76 & 1.24 & 16.9 & 0.23 & 2.77 & 0.0 & -- & No \\
\midrule
         40.0 & Single & 5.4 & 30.7 & 23.3 & 15.8 & 2.45 & 0.15 & 0.77 & 1.28 & 30.7 & 0.41 & 2.43 & 0.0 & No & No \\
         40.0 & \casea & 5.5 & 16.7 & 16.7 & 14.1 & 2.07 & 0.19 & 0.66 & 1.17 & 16.7 & 0.19 & 2.45 & 0.0 & -- & No \\
         40.0 & \casebe & 5.4 & 16.0 & 16.0 & 13.5 & 2.00 & 0.20 & 0.59 & 1.14 & 16.0 & 0.17 & 2.35 & 0.0 & -- & No \\
         40.0 & \casebl & 5.4 & 16.0 & 16.0 & 13.6 & 1.92 & 0.20 & 0.41 & 1.09 & 1.77 & 0.12 & 2.11 & 14.0 & -- & No \\
         40.0 & \casec & 5.4 & 18.6 & 18.6 & 15.8 & 2.13 & 0.15 & 0.61 & 1.16 & 18.6 & 0.34 & 1.84 & 0.0 & -- & No \\
\midrule
         42.0 & Single & 5.2 & 32.2 & 25.2 & 17.0 & 1.79 & 0.14 & 0.39 & 1.03 & 32.2 & 0.13 & 1.88 & 0.0 & No & No \\
         42.0 & \casea & 5.3 & 18.1 & 18.1 & 15.4 & 2.24 & 0.19 & 0.71 & 1.19 & 18.1 & 0.31 & 2.40 & 0.0 & -- & No \\
         42.0 & \casebe & 5.3 & 17.0 & 17.0 & 14.5 & 2.13 & 0.19 & 0.68 & 1.17 & 17.0 & 0.28 & 2.38 & 0.0 & -- & No \\
         42.0 & \casebl & 5.3 & 17.1 & 17.1 & 14.6 & 2.09 & 0.19 & 0.66 & 1.15 & 17.1 & 0.29 & 2.27 & 0.0 & -- & No \\
         42.0 & \casec & 5.2 & 19.9 & 19.9 & 17.0 & 1.95 & 0.14 & 0.42 & 1.08 & 19.9 & 0.13 & 2.05 & 0.0 & -- & No \\
\midrule
         45.0 & Single & 5.0 & 34.9 & 26.3 & 18.5 & 3.17 & 0.14 & 0.62 & 1.19 & 34.9 & 0.28 & 1.80 & 0.0 & No & No \\
         45.0 & \casea & 5.1 & 19.6 & 19.6 & 16.8 & 2.36 & 0.18 & 0.76 & 1.24 & 19.6 & 0.35 & 2.49 & 0.0 & -- & No \\
         45.0 & \casebe & 5.0 & 18.7 & 18.7 & 16.1 & 2.25 & 0.18 & 0.73 & 1.22 & 18.7 & 0.34 & 2.41 & 0.0 & -- & No \\
         45.0 & \casebl & 5.0 & 18.7 & 18.7 & 16.2 & 2.22 & 0.18 & 0.68 & 1.18 & 18.7 & 0.33 & 2.22 & 0.0 & -- & No \\
         45.0 & \casec & 5.0 & 21.5 & 21.5 & 18.5 & 2.01 & 0.14 & 0.46 & 1.13 & 21.5 & 0.13 & 2.16 & 0.0 & -- & No \\
\midrule
         47.0 & Single & 4.9 & 36.0 & 28.8 & 19.9 & 2.79 & 0.13 & 0.79 & 1.41 & 36.0 & 0.52 & 1.69 & 0.0 & No & No \\
         47.0 & \casea & 4.9 & 20.7 & 20.7 & 17.8 & 2.16 & 0.17 & 0.59 & 1.16 & 20.7 & 0.32 & 1.87 & 0.0 & -- & No \\
         47.0 & \casebe & 4.9 & 19.5 & 19.5 & 16.9 & 2.31 & 0.18 & 0.71 & 1.23 & 19.5 & 0.36 & 2.28 & 0.0 & -- & No \\
         47.0 & \casebl & 4.9 & 19.8 & 19.8 & 17.2 & 2.10 & 0.18 & 0.56 & 1.13 & 19.8 & 0.26 & 1.98 & 0.0 & -- & No \\
         47.0 & \casec & 4.9 & 22.9 & 22.9 & 19.9 & 2.04 & 0.13 & 0.50 & 1.14 & 22.9 & 0.14 & 2.20 & 0.0 & -- & No \\
\midrule
         50.0 & Single & 4.7 & 38.2 & 27.0 & 21.9 & 2.03 & 0.14 & 0.45 & 1.11 & 38.2 & 0.13 & 2.12 & 0.0 & No & No \\
         50.0 & \casea & 4.7 & 22.5 & 22.5 & 19.6 & 2.33 & 0.17 & 0.65 & 1.21 & 22.5 & 0.37 & 1.74 & 0.0 & -- & No \\
         50.0 & \casebe & 4.7 & 21.3 & 21.3 & 18.7 & 2.27 & 0.17 & 0.68 & 1.21 & 21.3 & 0.38 & 1.81 & 0.0 & -- & No \\
         50.0 & \casebl & 4.7 & 21.6 & 21.6 & 19.0 & 2.42 & 0.17 & 0.80 & 1.30 & 21.6 & 0.21 & 2.88 & 0.0 & -- & No \\
         50.0 & \casec & 4.7 & 24.8 & 24.8 & 21.9 & 2.20 & 0.14 & 0.52 & 1.15 & 24.8 & 0.14 & 2.25 & 0.0 & -- & No \\
\midrule
         55.0 & Single & 4.4 & 43.6 & 35.4 & 24.4 & 2.27 & 0.12 & 0.65 & 1.25 & 43.6 & 0.21 & 2.26 & 0.0 & No & No \\
         55.0 & \casea & 4.5 & 25.2 & 25.2 & 22.1 & 2.66 & 0.16 & 0.76 & 1.35 & 25.2 & 0.45 & 1.75 & 0.0 & -- & No \\
         55.0 & \casebe & 4.5 & 23.9 & 23.9 & 21.1 & 1.84 & 0.16 & 0.40 & 1.06 & 23.9 & 0.13 & 1.93 & 0.0 & -- & No \\
         55.0 & \casebl & 4.5 & 23.9 & 23.9 & 21.2 & 1.87 & 0.16 & 0.44 & 1.06 & 23.9 & 0.15 & 1.92 & 0.0 & -- & No \\
         55.0 & \casec & 4.4 & 27.7 & 27.7 & 24.4 & 2.29 & 0.12 & 0.67 & 1.24 & 27.7 & 0.18 & 2.39 & 0.0 & -- & No \\
\midrule
         60.0 & Single & 4.2 & 47.8 & 33.4 & 28.0 & 6.87 & 0.12 & 0.51 & 1.19 & 47.8 & 0.15 & 2.22 & 0.0 & No & No \\
         60.0 & \casea & 4.3 & 28.4 & 28.4 & 25.1 & 2.18 & 0.15 & 0.59 & 1.19 & 28.4 & 0.15 & 2.33 & 0.0 & -- & No \\
         60.0 & \casebe & 4.2 & 26.9 & 26.9 & 24.1 & 2.41 & 0.15 & 0.63 & 1.23 & 26.9 & 0.25 & 1.98 & 0.0 & -- & No \\
         60.0 & \casebl & 4.2 & 27.1 & 27.1 & 24.3 & 2.14 & 0.15 & 0.64 & 1.22 & 27.1 & 0.15 & 2.43 & 0.0 & -- & No \\
         60.0 & \casec & 4.2 & 31.2 & 31.2 & 28.0 & 2.27 & 0.12 & 0.80 & 1.32 & 31.2 & 0.20 & 2.60 & 0.0 & -- & No \\
\midrule
         65.0 & Single & 4.1 & 50.9 & 36.2 & 30.8 & 2.24 & 0.12 & 0.70 & 1.25 & 50.9 & 0.18 & 2.50 & 0.0 & No & No \\
         65.0 & \casea & 4.1 & 31.3 & 31.3 & 27.8 & 2.48 & 0.14 & 0.72 & 1.23 & 31.3 & 0.17 & 2.47 & 0.0 & -- & No \\
         65.0 & \casebe & 4.1 & 29.6 & 29.6 & 26.7 & 2.30 & 0.15 & 0.71 & 1.24 & 29.6 & 0.16 & 2.50 & 0.0 & -- & No \\
         65.0 & \casebl & 4.1 & 29.7 & 29.7 & 26.8 & 2.25 & 0.15 & 0.76 & 1.29 & 29.7 & 0.17 & 2.56 & 0.0 & -- & No \\
         65.0 & \casec & 4.1 & 34.0 & 34.0 & 30.8 & 2.71 & 0.12 & 0.81 & 1.35 & 34.0 & 0.26 & 2.82 & 0.0 & -- & No \\
\midrule
         70.0 & Single & 3.9 & 54.1 & 40.3 & 33.6 & 2.64 & 0.11 & 0.81 & 1.33 & 54.1 & 0.19 & 2.82 & 0.0 & No & No \\
         70.0 & \casea & 3.9 & 34.0 & 34.0 & 30.4 & 2.64 & 0.14 & 0.82 & 1.34 & 34.0 & 0.20 & 2.75 & 0.0 & -- & No \\
         70.0 & \casebe & 3.9 & 32.5 & 32.5 & 29.5 & 2.60 & 0.14 & 0.83 & 1.34 & 32.5 & 0.20 & 2.70 & 0.0 & -- & No \\
         70.0 & \casebl & 3.9 & 32.6 & 32.6 & 29.8 & 2.58 & 0.14 & 0.89 & 1.48 & 32.6 & 0.29 & 2.65 & 0.0 & -- & No \\
         70.0 & \casec & -- & -- & -- & -- & -- & -- & -- & -- & -- & -- & -- & -- & -- & -- \\
\midrule
         75.0 & Single & -- & -- & -- & -- & -- & -- & -- & -- & -- & -- & -- & -- & -- & -- \\
         75.0 & \casea & -- & -- & -- & -- & -- & -- & -- & -- & -- & -- & -- & -- & -- & -- \\
         75.0 & \casebe & 3.8 & 35.3 & 35.3 & 32.2 & 2.66 & 0.14 & 0.80 & 1.34 & 35.3 & 0.20 & 2.92 & 0.0 & -- & No \\
         75.0 & \casebl & 3.8 & 35.4 & 35.4 & 32.6 & 2.67 & 0.13 & 0.86 & 1.42 & 35.4 & 0.25 & 2.95 & 0.0 & -- & No \\
         75.0 & \casec & -- & -- & -- & -- & -- & -- & -- & -- & -- & -- & -- & -- & -- & -- \\
\midrule
         85.0 & Single & -- & -- & -- & -- & -- & -- & -- & -- & -- & -- & -- & -- & -- & -- \\
         85.0 & \casea & -- & -- & -- & -- & -- & -- & -- & -- & -- & -- & -- & -- & -- & -- \\
         85.0 & \casebe & -- & -- & -- & -- & -- & -- & -- & -- & -- & -- & -- & -- & -- & -- \\
         85.0 & \casebl & -- & -- & -- & -- & -- & -- & -- & -- & -- & -- & -- & -- & -- & -- \\
         85.0 & \casec & 3.6 & 46.4 & 46.4 & 42.0 & 2.37 & 0.10 & 0.74 & 1.25 & 46.4 & 0.36 & 2.42 & 0.0 & -- & No \\
\midrule
         90.0 & Single & -- & -- & -- & -- & -- & -- & -- & -- & -- & -- & -- & -- & -- & -- \\
         90.0 & \casea & 3.5 & 41.0 & 41.0 & 37.1 & 2.53 & 0.12 & 0.84 & 1.44 & 41.0 & 0.41 & 2.67 & 0.0 & -- & No \\
         90.0 & \casebe & 3.5 & 43.6 & 43.6 & 40.2 & 3.03 & 0.12 & 0.92 & 1.72 & 43.6 & 0.64 & 2.31 & 0.0 & -- & No \\
         90.0 & \casebl & -- & -- & -- & -- & -- & -- & -- & -- & -- & -- & -- & -- & -- & -- \\
         90.0 & \casec & -- & -- & -- & -- & -- & -- & -- & -- & -- & -- & -- & -- & -- & -- \\
    \end{longtable}
\end{ThreePartTable}
\twocolumngrid

\section{\label{app:bse-suppression}Suppression of \texorpdfstring{\bhl}{BHL}{+}\texorpdfstring{\bhh}{BHH} mergers by isolated binary star evolution}

We consider isolated binary-star evolution through the common-envelope and stable mass-transfer channels as illustrated in Fig.~\ref{fig:schematic-binary-evolution}. The evolution starts with two massive stars of initial masses $M_\mathrm{ini,1}$ and $M_\mathrm{ini,2}$ where $M_\mathrm{ini,1}>M_\mathrm{ini,2}$. The first mass-transfer phase has to be stable to form a BBH merger \citep{2016Natur.534..512B, 2021hgwa.bookE..16M}. The accretor is usually a main-sequence star that rejuvenates, \ie that adapts its interior structure to the new total mass $M_2$ after accretion. The first mass-transfer phase ends after the envelope mass $M_\mathrm{env,1}$ of star~1 has been transferred to star~2. After envelope loss, star~1 produces a BH from a \caseabh binary-stripped star. The second mass-transfer phase is either stable or leads to a common-envelope episode. In both cases, star~2 is also stripped of its envelope and produces another \BSS BH. The first-formed BH is usually assumed to not accrete significantly during the second mass-transfer phase because of Eddington-limited accretion.

Binary BH mergers invoking low-mass \bhl and high-mass \bhh are suppressed by the isolated binary-star evolution described above as shown as follows. Let $M_\mathrm{ini,1}\approx35\,\msun$ ($M_\mathrm{ini,1}\approx25\,\msun$) at $Z=\zsun$ ($Z=\zsun/10$) such that star~1 produces a low-mass \bhl because it falls into the compactness peak (Fig.~\ref{fig:compact-remnant-masses}; masses/values in parenthesis are for the lower metallicity models). In our models, star~1 has an envelope mass of $M_\mathrm{env,1}\approx17\,\msun$ ($M_\mathrm{env,1}\approx13\,\msun$) such that the accretor can at most reach a mass of $M_\mathrm{2,max}\approx 35\,\msun + 17\,\msun =52\,\msun$ if mass transfer is fully conservative ($M_\mathrm{2,max}\approx 25\,\msun + 13\,\msun =38\,\msun$). Such massive stars cannot experience \casec mass transfer in our models (Tables~\ref{tab:models_z} and~\ref{tab:models_z10}) and the second mass transfer thus leads to \BSSs that do not produce high-mass \bhh\ --- our models either predict NS or \bhl formation (Fig.~\ref{fig:compact-remnant-masses}). Hence, first forming \bhl in \bhl{+}\bhh mergers is forbidden in our models at both metallicities. 

Next, we consider that the more massive \bhh forms first. This requires initial masses of $M_\mathrm{ini,1}>70\,\msun$ ($M_\mathrm{ini,1}>40\,\msun$) because star~1 will otherwise produce a NS or a low-mass \bhl (Fig.~\ref{fig:compact-remnant-masses}). As before, we assume that the initial masses of the secondary star, $M_{\mathrm{ini},2}$, are drawn from a uniform mass ratio distribution, \ie all $M_{\mathrm{ini},2}$ are equally likely for a given $M_{\mathrm{ini},1}$. For an arbitrary, fixed mass-transfer efficiency, there is only a very limited range of initial masses for star~2, $\Delta M$, ($M_\mathrm{ini,2}<M_\mathrm{ini,1}$) to obtain a mass of $M_2\approx35\,\msun\pm\Delta M$ ($M_2\approx25\,\msun\pm\Delta M$) after the first mass-transfer phase such that star~2 falls into the compactness peak and then leaves behind a low-mass \BSS \bhl. In the majority of cases, star~2 has a different mass and does not fall into the compactness peak such that it forms a \BSS NS or \bhh after the second mass-transfer phase. Let $M_\mathrm{min}$ and $M_\mathrm{max}$ be the minimum and maximum (initial) mass of a star to form a NS or BH, respectively, and $M_{\mathrm{BH}_\mathrm{H}}$ the (initial) mass threshold for \bhh formation. The probability of forming a compactness-peak \bhl after the second mass transfer is then $\Delta M/(M_\mathrm{max}-M_\mathrm{min})$ while the probabilities to form a NS and \bhh are $(M_{\mathrm{BH}_\mathrm{H}} - M_\mathrm{min} - \Delta M)/(M_\mathrm{max}-M_\mathrm{min})$ and $(M_\mathrm{max} - M_{\mathrm{BH}_\mathrm{H}})/(M_\mathrm{max}-M_\mathrm{min})$, respectively. \bhl{+}\bhh mergers where \bhh forms first are thus suppressed compared to NS{+}\bhh and \bhh{+}\bhh mergers by factors of $(M_{\mathrm{BH}_\mathrm{H}} - M_\mathrm{min} - \Delta M)/\Delta M$ and $(M_\mathrm{max} - M_{\mathrm{BH}_\mathrm{H}})/\Delta M$, respectively. Assuming $M_\mathrm{min}\approx10\,\msun$, $M_\mathrm{max}\approx100\,\msun$, $\Delta M \approx 5\,\msun$ and taking $M_{\mathrm{BH}_\mathrm{H}}\approx70\,\msun$ at $\zsun$ and $\approx40\,\msun$ at $\zsun/10$, we find suppression factors of about $5\text{--}10$. We conclude that \bhl{+}\bhh mergers are not forbidden but suppressed in our models, and the models predict that some BBHs in the $10\text{--}12\,\msun$ chirp-mass gap should be detected in the future.

\section{\label{app:influence-bhs-physics}Influence of various physics assumptions on the BH mass spectrum}

\begin{figure}
    \centering
    \includegraphics{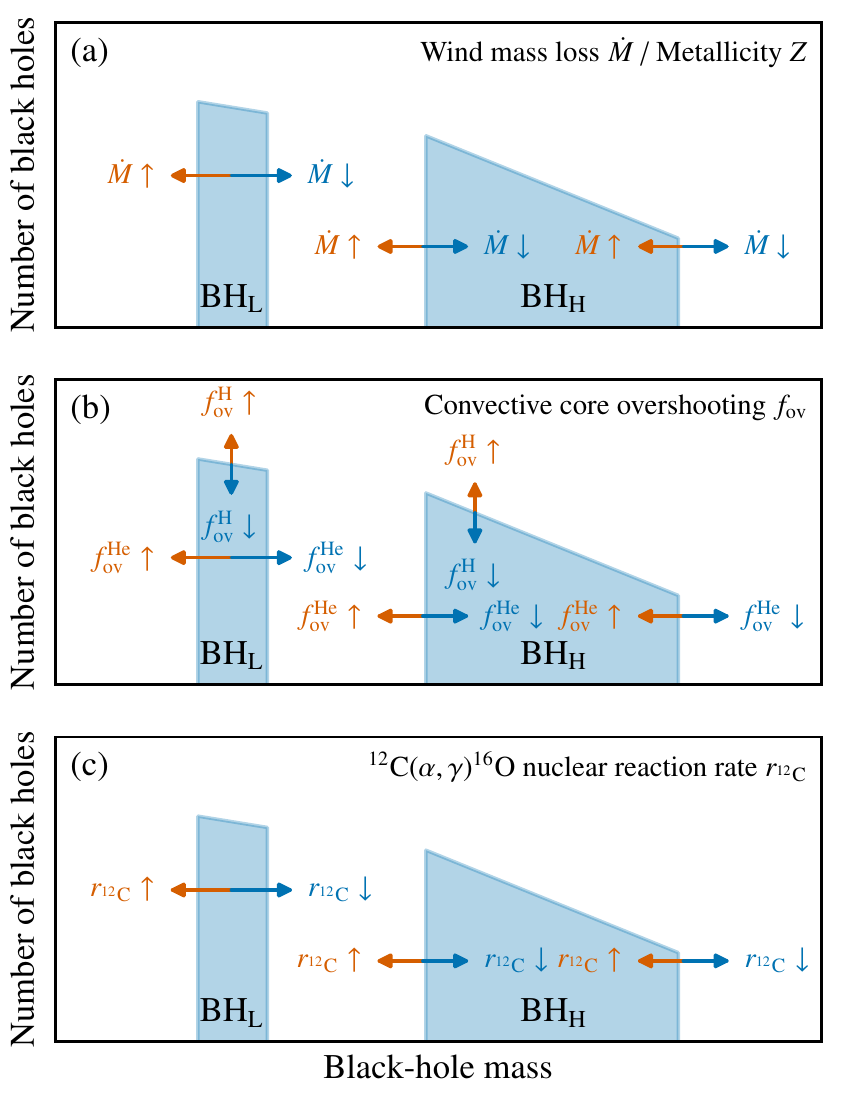}
    \caption{\label{fig:schematic-bh-mass-distribution}Influence of selected physical processes on the BH mass spectrum of \BSSs. Illustrated are systematic variations by an increase/decrease of the wind mass-loss rate $\dot{M}$ (\eg via a larger/smaller metallicity $Z$; panel a), of the convective core boundary mixing during core hydrogen and core helium burning ($f_\mathrm{ov}^{\mathrm{H}}$ and $f_\mathrm{ov}^{\mathrm{He}}$, respectively; panel b) and of the $^{12}\mathrm{C}(\alpha,\gamma)^{16}\mathrm{O}$ nuclear reaction rate (panel c). Exact quantitative differences are currently unknown and the subject of future research.}
\end{figure}

The exact bimodal structure of the BH mass spectrum of single and \BSSs depends on various physical processes and as such offers the possibility to better understand and constrain them with the help of observations. In Fig.~\ref{fig:schematic-bh-mass-distribution}, we schematically show how the masses of the low-mass \bhl and high-mass \bhh may change, and how the overall number of BHs in comparison to NSs may vary for changes in the wind mass-loss rates of stars, convective boundary mixing and the $^{12}\mathrm{C}(\alpha,\gamma)^{16}\mathrm{O}$ nuclear reaction rate. There are further physical processes and other uncertain ingredients in stellar models that may affect the BH mass spectrum but we limit our discussion to the aforementioned aspects.

Wind mass-loss rates $\dot{M}$ are still considerably uncertain and could in reality be lower or higher compared to what is assumed in our models \citep{2014ARA&A..52..487S}. Similarly, mass-loss rates of stars are higher at higher metallicity and lower at smaller metallicity. The indicated changes in Fig.~\ref{fig:schematic-bh-mass-distribution}a are thus meant to cover systematic uncertainties in $\dot{M}$ and stars at different metallicities. In this work, we have shown that \bhl and the lower end of \bhh masses depend only mildly on $\dot{M}$, and that they are systematically smaller for higher $\dot{M}$ and thus larger metallicity (and vice versa). In contrast, the maximum BH mass depends strongly on the wind strength and metallicity, with BH masses being larger at lower $\dot{M}$ and smaller metallicity, and vice versa \citep{2005A&A...442..587V, 2010ApJ...714.1217B, 2015MNRAS.451.4086S}. 

In our models, we consider extra mixing at the boundary of convective cores during core hydrogen and core helium burning. This extra mixing is parametrized as step overshooting by an amount of $f_\mathrm{ov}$ as measured in units of the local pressure scale height. In \BSSs, \ie in helium stars, the overshooting during core helium burning sets the size of the resulting CO-core mass and thus the ratio of total helium-star to CO-core mass. Larger overshooting then implies a smaller total and hence smaller BH mass for the characteristic CO-core masses that result in low-mass \bhl and the lowest-mass \bhh (Fig.~\ref{fig:schematic-bh-mass-distribution}b). For the highest-mass BHs, larger overshooting during both core hydrogen and even more so during core helium burning imply larger stellar luminosity for a given total mass, hence higher wind mass loss and smaller BH masses.

Step overshooting during core hydrogen burning sets the helium-core masses of stars. It thus links the helium masses of \BSSs to their initial mass and thus to their overall fraction as given by the stellar initial mass function. A larger overshooting value means that initially less massive stars can produce the same helium-star mass; hence, the overall fraction of BHs increases with larger overshooting during core hydrogen burning (Fig.~\ref{fig:schematic-bh-mass-distribution}b). In \BSSs, overshooting during core hydrogen burning does not directly affect the resulting \bhl and \bhh masses as they are rather connected to the CO-core masses set by core helium burning.

The $^{12}\mathrm{C}(\alpha,\gamma)^{16}\mathrm{O}$ nuclear reaction rate $r_{^{12}\mathrm{C}}$ governs the abundance of $^{12}\mathrm{C}$ remaining at the end of core helium burning \citep{1989A&A...210...93L, 1993ApJ...411..823W, 2001NewA....6..457B}. A faster reaction rate converts more carbon into oxygen and vice versa. Fewer carbon atoms during core carbon burning mean that neutrinos overcome energy generation from nuclear burning already at lower CO-core masses and so the compactness peak systematically shifts to lower CO-core masses. This in turn results in less massive compactness-peak BHs (Fig.~\ref{fig:schematic-bh-mass-distribution}c). The abundance of neon after core carbon burning is less if there are fewer carbon atoms after helium burning. Hence, also the increase of compactness related to neutrinos overcoming the energy generation from core neon burning is found at lower CO-core masses such that the corresponding BH masses are smaller.

The maximum BH mass shown in Fig.~\ref{fig:schematic-bh-mass-distribution} is connected to wind mass loss and the occurrence of pair-instability supernovae (PISNe). Assuming that it is set by PISNe that leave no BH remnants, we indicate in Fig.~\ref{fig:schematic-bh-mass-distribution} the finding of \citep{2020ApJ...902L..36F} that the lower end of the PISN BH-mass gap is at higher masses for a slower $r_{^{12}\mathrm{C}}$ and vice versa. The masses of BHs beyond the PISN mass gap are not shown.

Quantitatively, decreasing $Z$ by a factor of 10 as done in this work, \ie decreasing $\dot{M}$ by about a factor of 3, decreases $M_\mathrm{CO}$ of the compactness peak by $0.5\,\msun$ in single stars and $0.8\,\msun$ in \BSSs. In the \BSSs, this translates into an increase of the mass of compactness-peak BHs by a similar amount (${\approx}\,0.6\,\msun$).

Increasing and decreasing the overshooting parameter in our single-star models by factors of 2, we find that the compactness-peak $M_\mathrm{CO}$ increases by ${\approx}\,0.5\,\msun$ and decreases by ${\approx}\,0.3\,\msun$, respectively. As for the change in $Z$, we expect a similar change in the mass of compactness-peak BHs of \BSSs.

When using the nuclear reaction rate $r_{^{12}\mathrm{C}}$ of \citep{2002ApJ...567..643K}, which is about 20\% slower than the one used in our models \citep{2013NuPhA.918...61X}, the $M_\mathrm{CO}$ of the compactness peak gets ${\approx}\,0.9\,\msun$ larger. Similarly, boosting $r_{^{12}\mathrm{C}}$ in our models by 10\%, leads to ${\approx}\,0.4\,\msun$ smaller compactness-peak $M_\mathrm{CO}$. Again, the corresponding BH masses of \BSSs are expected to change by a similar amount of mass.


\bibliographystyle{aasjournal}
\newcommand{\noop}[1]{}



\end{document}